\documentclass[11 pt]{article}
\usepackage[margin = 0.75 in]{geometry}
\usepackage{amsmath}
\usepackage{float}
\usepackage{graphicx}
\usepackage{authblk}
\usepackage{braket}
\usepackage{cite}

\providecommand{\keywords}[1]
{
  \small	
  \textbf{\textit{Keywords---}} #1
}

\title{Floquet–Bloch Theory for Dispersive Time-Varying Metasurfaces}

\author[1]{Mohammad Mojtaba Sadafi}
\author[1]{Sandeep Inampudi}
\author[1]{Hossein Mosallaei \thanks{corresponding author, email: h.mosallaei@northeastern.edu}}

\affil[1]{Metamaterials Laboratory, Electrical and Computer Engineering Department, Northeastern University, Boston, MA 02115, USA}

\date{}

\begin{document}

\maketitle

\begin{abstract}
Light–matter interaction in time-varying metasurfaces brings about phenomena that transcend the limits of static systems. Temporal modulation enables energy exchange between light and matter, and electromagnetic fields are coupled in both momentum and frequency, with implications across a broad range of photonic applications. Nevertheless, a precise description of such systems necessitates a theory that captures the intertwined effects of spatiotemporal variations while accounting for dispersion in a causal manner, as realistic optical materials exhibit frequency-dependent response and finite temporal memory. Here, a self-contained Floquet–Bloch theory is developed to capture the response of dispersive, time-varying metasurfaces, respecting causality through physically consistent constitutive relations. The theory treats spatial periodicity, nonadiabatic temporal modulation, and material dispersion within a unified formalism, providing access not only to the metasurface's scattering response but also to its inherent modal structure. The formulation is validated against full-wave simulations. As illustrative examples, it is first applied to asymmetric Floquet harmonic generation in an excitonic time-varying metasurface, where excitonic dispersion enables selective harmonic enhancement. It is then used to investigate metasurface-based photonic time crystals, revealing how modal dispersion governs momentum-bandgap formation and dynamics. This work establishes a comprehensive platform for understanding dispersive, time-varying metasurfaces and their underlying physical mechanisms.
\end{abstract}


\newpage
\section{Introduction}
Temporal variations, as an additional degree of freedom, can be introduced into optical materials to enable light--matter interaction phenomena that are inherently inaccessible to static media \cite{galiffi2022photonics,patel2026photonic,engheta2023four,asgari2024theory}. As a direct result of broken temporal translation symmetry, the frequency of light is no longer preserved upon interacting with time-varying media, allowing energy to be exchanged between the electromagnetic fields and the temporal modulation process \cite{galiffi2025electrodynamics,lyubarov2022amplified,galiffi2026optical}. Together, these mechanisms give rise to a new class of optical effects, ranging from frequency conversion and nonreciprocity to parametric amplification and temporal refraction \cite{salary2019dynamically,shaltout2015time,tirole2023double,koutserimpas2018parametric,dong2024quantum}. In this context, time-varying photonic structures have opened up a broad range of applications across diverse fields, including dynamic directional light control, magnetless optical isolation, enhanced wireless power transfer, quantum state engineering, topological light control, and photonic time crystals (PTCs) \cite{sadafi2023dynamic,asadchy2022parametric,estep2014magnetic,barati2022optical,wang2024time,sadafi2025time,da2025dynamics,horsley2023quantum,zhu2022time,xiong2025observation,dikopoltsev2022light,li2023stationary}. Through a broader lens, spatial geometry and resonant response of the medium can play a substantial role by enhancing light--matter interaction, thereby strengthening the influence of time modulation on the optical response of the system.\\
Among various optical platforms, metamaterials as well as their planar counterparts, metasurfaces, hold immense potential when endowed with periodic nonadiabatic temporal modulation at rates comparable to their operating frequency, as they offer a route toward manipulating both the momentum and energy of light through the combined control of spatial and temporal degrees of freedom \cite{sabri2023high,barati2020topological,wu2025space,hu2022arbitrary,garg2025inverse}. In time-varying metasurfaces, spatial periodicity and temporal modulation are intertwined, and optical fields are coupled across both momentum and frequency. This, in turn, projects the electromagnetic response onto a discrete set of Floquet--Bloch waves that inherit their attributes not only from the inherent geometrical properties of the metasurface, but also directly from its temporal modulation profile, which opens up an avenue toward controlling scattering, dispersion, and modal interactions \cite{barati2020time,wang2025expanding,ma2025floquet}. As an illustration, spatiotemporal modulation of metasurfaces has been utilized to generate optical vortices with time-varying orbital angular momentum and topological charge \cite{barati2020time}, while resonant Mie-type metasurfaces have been shown to enhance momentum bandgaps in PTCs \cite{wang2025expanding}. More recently, programmable time-varying metasurfaces have also enabled Floquet topological phases of light, in which periodic temporal modulation breaks time-reversal symmetry and supports chiral edge states without resorting to magneto-optical materials \cite{ma2025floquet}. Nevertheless, realizing such applications is demanding in practice, as it requires materials whose optical response can be tuned strongly and at high speeds.\\ 
Various material platforms have been explored to realize dynamic metasurfaces, ranging from transparent conducting oxides such as indium tin oxide (ITO) operated in the epsilon-near-zero (ENZ) regime, to excitonic van der Waals materials such as transition metal dichalcogenides (TMDCs) and aligned carbon nanotubes (CNTs), as well as nonlinear electro-optic crystals such as lithium niobate (LN) and barium titanate (BTO) \cite{sisler2024electrically,jaffray2026all,liu2021photon,dhama2026cross,kumar2021light,lynch2025electrically,di2025efficient,wang2018integrated}. Obtaining large modulation strengths in these platforms often necessitates operating near material resonances, electronic transitions, or band-edge features, where the optical response changes rapidly with frequency and becomes strongly dispersive \cite{lynch2025full,sadafi2021tunable,khurgin2024energy}. Consequently, a dispersionless treatment of the medium, while convenient and widely adopted in various previous works, is unable to capture the harmonic asymmetries, modal couplings, and resonance-enhanced momentum bandgap structures that arise when each Floquet harmonic samples a different region of the material frequency response. Therefore, a self-consistent theory for light--matter interaction in dispersive, time-varying materials is crucial for accurately describing realistic dynamic metasurfaces.\\
Proper incorporation of dispersion into the theoretical treatment of nonadiabatic time-varying metasurfaces is not only essential for capturing the frequency-dependent behavior of different Floquet harmonics, but also necessary for describing light--matter interaction in causal optical materials in a physically consistent manner \cite{hayran2022homega,sloan2024optical,koutserimpas2024time}. Nonetheless, most existing approaches rely either on adiabatic or dispersionless descriptions of temporal modulation, which hold only for small modulation frequencies or become physically incomplete from a more general perspective, since the assumption of an instantaneous material response neglects the finite temporal memory of any real medium \cite{salary2018time,inampudi2018rigorous,taravati2017nonreciprocal,stefanou2023light}. In this sense, different approaches have recently been developed to address the interplay between material dispersion and temporal modulation \cite{mirmoosa2022dipole,ptitcyn2023floquet,garg2022modeling}. At the level of subwavelength scatterers, the polarizability of time-varying particles has been reformulated in a nonstationary and causal fashion, introducing temporal polarizability kernels that simultaneously account for both memory and explicit time dependence of the material response. This perspective, in turn, provides insight into the underlying physics of dispersive, time-varying meta-atoms and their effective material response \cite{mirmoosa2022dipole}. Within this framework, Floquet--Mie theory generalizes the classical Mie formalism to dispersive spheres with periodically modulated material properties by expanding the fields in terms of vector spherical harmonics (VSHs) and deriving the corresponding Floquet T-matrix, which encapsulates light--matter interaction with the particle \cite{ptitcyn2023floquet}. Furthermore, this T-matrix description has been extended to periodic arrangements of time-varying scatterers, where the response of the metasurface is constructed from the single-particle T-matrix together with the translation coefficients and lattice sums associated with the vector spherical wave basis in a two-dimensional periodic lattice \cite{garg2022modeling}. These developments, together with other recent contributions, have established powerful analytical and semi-analytical tools for modeling dispersive time-varying photonic systems \cite{sun2025formulation,de2025lattice,verde2026optical,iplikcciouglu2025analytical,movahediqomi2026stacked}. The remaining gap lies in a self-contained theoretical framework for dispersive time-varying metasurfaces, particularly in the nonadiabatic regime where the modulation speed is comparable to the operating frequency, in which spatial periodicity and temporal modulation are treated consistently within a coupled Floquet--Bloch description, and from which modal dispersion, scattering response, and power exchange between the field and the modulating medium can be derived within a comprehensive formulation.\\
In this work, we develop a self-consistent Floquet--Bloch theory for light--matter interaction in dispersive, nonadiabatic time-varying metasurfaces, in which causality is incorporated at the constitutive level through causal material response kernels consistent with the Kramers--Kronig relations. First, by treating spatial periodicity and temporal modulation as intertwined degrees of freedom, we project the electromagnetic response onto a Floquet–Bloch basis and couple the spatiotemporal harmonics through the frequency-dependent susceptibility kernel of the medium that encodes dispersion. This, in turn, allows us to track how each harmonic samples a distinct region of the material response and to unravel the role of dispersion in harmonic coupling and modal formation. Next, we expand the electromagnetic fields within each layer of a finite metasurface stack in this coupled basis to derive a modal eigenvalue problem that describes the dynamics of the Floquet–Bloch waves, and resort to a scattering-matrix formalism to impose the boundary conditions between the layers and obtain the collective response of the structure. In this sense, the implementation is dual to the rigorous coupled-wave analysis (RCWA) used for conventional metasurfaces, which provides an exceptional tool for studying passive multilayer systems \cite{moharam1981rigorous,moharam1995stable}. Finally, we formulate a rigorous Floquet–Bloch Poynting theorem in order to connect the modal picture with the power carried by individual Floquet channels, which provides access to the observable scattering characteristics of the metasurface. Our theory not only addresses externally driven light–matter interaction, but also resolves the quasinormal-mode spectrum and dispersion of the time-varying metasurface, allowing us to study intrinsic phenomena such as the modal physics of PTCs directly within the same framework.\\
We validate our formalism against full-wave finite-element simulations and demonstrate that the dispersive Floquet--Bloch theory accurately captures the harmonic spectra and field distributions of time-modulated dielectric gratings in regimes where dispersionless models are no longer adequate. We then apply the theory to two representative settings that highlight its physical reach. First, we study asymmetric Floquet harmonic generation in a time-modulated metasurface, in which the highly dispersive response of a monolayer excitonic material, combined with a resonant guided mode of a dielectric grating, breaks the symmetry between up- and down-converted harmonics, resulting in the selective excitation of the desired sideband even under a simple sinusoidal modulation. Our analysis demonstrates that this effect is inherently tied to material dispersion and therefore cannot be captured within a dispersionless model of the system. Second, we apply our theory to a time-varying metasurface and analyze its inherent modal structure as a PTC. We show that the quasinormal-mode branches of the static metasurface are coupled through temporal modulation, giving rise to Floquet--Bloch bands with avoided crossings, momentum bandgaps, and complex eigenfrequencies. We further show that the dispersion of the underlying Floquet--Bloch modes can be harnessed to control the size of the momentum bandgap, opening a route toward PTCs based on time-varying metasurfaces under weak temporal modulation. Our study establishes dispersive Floquet--Bloch theory as a comprehensive approach to understanding the underlying physics of dispersive spatiotemporal light--matter interaction under fast nonadiabatic temporal modulation, and provides a powerful framework for the analysis, design, and optimization of time-varying metasurfaces.\\
The rest of this paper is organized as follows. Section 2 builds the theoretical basis of our work by developing the dispersive Floquet--Bloch formalism, the scattering description of layered time-varying metasurfaces, and the corresponding analysis of the power flow. Section 3 introduces the causal dispersive material model and validates the theory against full-wave simulations. Section 4 applies the theory to asymmetric Floquet harmonic generation and metasurface-enabled PTCs. Finally, Section 5 summarizes the main findings and discusses their implications for dispersive spatiotemporal photonic platforms.
\section{Theoretical Framework}
In this section, we seek to develop a comprehensive theory to address light--matter interaction with metasurfaces composed of dispersive time-varying materials in the context of Floquet--Bloch waves. We start by characterizing the dynamics of electromagnetic waves in a bulk medium with spatiotemporally modulated optical properties in terms of an eigenvalue problem that incorporates dispersion by means of a frequency-dependent convolution operator. We further extend our formalism to capture the collective response of the metasurface stack by invoking the proper eigenmode expansion and relating the modal amplitudes through a scattering matrix. Finally, we define the notion of power flow in dispersive time-varying metasurfaces by formulating a Floquet--Bloch Poynting theorem, which provides a modal description of electromagnetic power transport and enables the evaluation of harmonic power redistribution and observable scattering characteristics.
\begin{figure}[t]
    \centering
    \includegraphics[scale = 1]{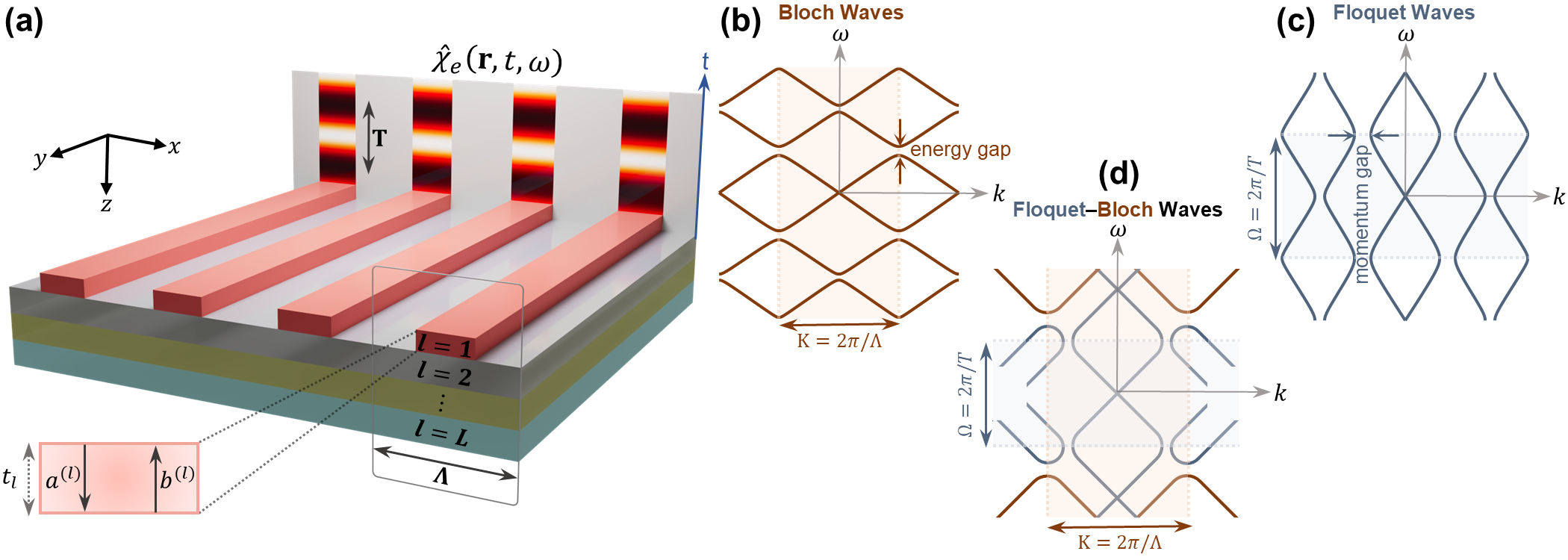}
    \caption{Conceptual framework for Floquet–Bloch theory of dispersive time-varying metasurfaces. (a) Multilayer metasurface consisting of a finite stack of spatiotemporally modulated dispersive layers arranged along the propagation direction. Each layer is defined by its thickness $t_l$ and susceptibility profile, with the field response described through forward and backward modal amplitudes $a^{(l)}$ and $b^{(l)}$. (b--d) Conceptual band diagrams comparing Bloch, Floquet, and Floquet--Bloch waves. Spatial harmonic coupling in Bloch media opens frequency gaps, while temporal harmonic coupling in Floquet media gives rise to momentum gaps. In spatiotemporally periodic media, both coupling mechanisms coexist, enabling simultaneous gap features and modulation-induced amplification.}
    \label{fig.1}
\end{figure}
\subsection{Floquet–Bloch Formalism for Spatiotemporally Periodic Dispersive Media}
We investigate the behavior of electromagnetic waves in a linear, nonmagnetic, and dispersive medium whose optical properties are modulated periodically in both space and time. For clarity, our study is restricted to two-dimensional geometries that exhibit spatial periodicity along one in-plane direction and invariance along the other (see Figure 1a), whereas the three-dimensional case can be addressed by a straightforward mathematical extension of the present analysis. In such a medium, the dynamics of the solutions are governed by the following form of Maxwell's equations
\begin{subequations}
\begin{equation}
    \nabla\times\textbf{E}(\textbf{r},t)=-\frac{\partial}{\partial t}\,\textbf{B}(\textbf{r},t),
\end{equation}
\begin{equation}
    \nabla\times\textbf{H}(\textbf{r},t)=\frac{\partial}{\partial t}\,\textbf{D}(\textbf{r},t),
\end{equation}
\end{subequations}
with $\textbf{D}=\varepsilon_0\textbf{E}+\textbf{P}$ and $\textbf{B}=\mu_0\textbf{H}$. Here, $\varepsilon_0$ and $\mu_0$ are the permittivity and permeability of vacuum, respectively, and $\mathbf{P}$ is the polarization vector, which describes the temporal response of the medium to the applied electric field. In a dispersive time-varying medium, it is expressed through the convolution integral \cite{mirmoosa2022dipole},
\begin{equation}
    \textbf{P}(\textbf{r},t)=\varepsilon_0\int_{-\infty}^{+\infty}\chi_e(\textbf{r},t,t-\tau)\,\mathbf{E}(\textbf{r},\tau)\,d\tau.
\end{equation}
In Equation (2), $\chi_e(\mathbf{r},t,\gamma)$ denotes the electric susceptibility of the medium, which accounts for spatial and temporal variations as well as dispersion by treating $t$ as the absolute observation time and $\gamma=t-\tau$ as the delay variable associated with the temporal memory of the medium. In contrast to the idealized assumption of an instantaneous material response, the polarization-based constitutive relation adopted here enables fast temporal modulation to be treated in arbitrarily dispersive media without neglecting the finite memory of the material. Causality is incorporated by requiring $\chi_e(\mathbf{r},t,\gamma)=0$ for $\gamma<0$, which leads to the corresponding Kramers--Kronig relations when the required regularity conditions are satisfied \cite{koutserimpas2024time, mirmoosa2022dipole}. Using the proper Fourier convention (see Supporting Information, Section 1) and following some mathematical simplifications, the frequency-domain polarization vector is given by
\begin{equation}
    \tilde{\textbf{P}}(\textbf{r},\omega)=\frac{\varepsilon_0}{2\pi}\int_{-\infty}^{+\infty}
    \tilde{\chi}_e(\textbf{r},\omega-\omega',\omega')\,\tilde{\mathbf{E}}(\mathbf{r},\omega')\,d\omega',
\end{equation}
which is used to obtain the system's frequency response and explicitly captures the frequency mixing induced by temporal modulation. Here and in the following, quantities with a tilde denote their frequency-domain counterparts, with $\tilde{\chi}_e$ obtained by Fourier transforming the temporal arguments of $\chi_e$.\\
Next, we assume that the susceptibility is periodic along $x$ with period $\Lambda$ and periodic in time with period $T$. The function $\hat{\chi}_e(x,t,\omega')$ can then be expanded as a spatiotemporal Fourier series, with the dependence on the delay variable $\gamma$ incorporated through its Fourier transform, as \cite{inampudi2018rigorous}
\begin{equation}
    \hat{\chi}_e(x,t,\omega')=\sum_{n,m}\hat{\chi}_{nm}(\omega')\,e^{imK{x}}\,e^{-in\Omega{t}},
\end{equation}
where $K=2\pi/\Lambda$ and $\Omega=2\pi/T$ denote the reciprocal lattice vector and temporal modulation angular frequency, respectively, and $\hat{\chi}_{nm}(\omega')$ is the corresponding spatiotemporal Fourier coefficient, accounting for the dispersive nature of the material (see Supporting Information, Section 2). Owing to the discrete translational symmetry of the system in both space and time, the electromagnetic fields can be expressed using the Floquet--Bloch expansion,
\begin{subequations}
\begin{align}
    \textbf{E}(x,z,t)&=e^{ik_x{x}}e^{-i\omega{t}}\sum_{n,m}\mathbf{E}_{nm}(z)\,e^{imK{x}}\,e^{-in\Omega{t}},\\
    \mathbf{H}(x,z,t)&=e^{ik_x{x}}e^{-i\omega{t}}\sum_{n,m}\mathbf{H}_{nm}(z)\,e^{imK{x}}e^{-in\Omega t},
\end{align}
\end{subequations}
in which $k_x\in[-K/2,K/2]$ is the Bloch wavevector in the first spatial Brillouin zone, $\omega\in[-\Omega/2,\Omega/2]$ is the Floquet quasifrequency in the first temporal Brillouin zone, and $\mathbf{E}_{nm}$ and $\mathbf{H}_{nm}$ denote the corresponding harmonic field components. Within the Floquet--Bloch representation, and using the convolution form of the dispersive susceptibility, the polarization harmonic evaluated at $\omega_n=\omega+n\Omega$ and $k_m=k_x+mK$ is governed by
\begin{equation}
    \tilde{\mathbf{P}}_{nm}(z)
    =
    \varepsilon_0
    \sum_{n',m'}
    \hat{\chi}_{n-n',\,m-m'}\!\left(\omega_{n'}\right)\,
    \mathbf{E}_{n'm'}(z),
\end{equation}
which establishes the dispersive Floquet--Bloch coupling rule of the system, where spatiotemporal harmonic mixing is mediated by a convolution kernel that depends explicitly on the source frequency. As a result, each coupling channel retains the material dispersion associated with the harmonic from which it originates.\\
Due to translational symmetry along the $z$-direction, the field solutions can be taken to have the longitudinal dependence $\exp\!\left(i\beta{z}\right)$, where $\beta$ is the propagation constant associated with a Floquet--Bloch eigenmode. Projecting the Floquet--Bloch representation onto a finite set of temporal and spatial harmonics, Maxwell's equations reduce to the dispersive Floquet--Bloch eigenvalue problem
\begin{equation}
    \mathbf{M}\boldsymbol{\Psi}_e
    =
    \boldsymbol{\Psi}_e\boldsymbol{\beta}^{2},
\end{equation}
where $\mathbf{M}$ is the dispersive Floquet--Bloch operator, $\boldsymbol{\Psi}_e = \left[\textbf{E}^{(x)},\,\textbf{E}^{(y)}\right]^{\mathrm{T}}$ collects the electric-field eigenvectors, and $\boldsymbol{\beta}=\operatorname{diag}\left[\beta_q\right]$ contains the corresponding propagation constants. The eigenvalue spectrum therefore yields the complex propagation constants $\beta_q(\omega,k_x)$ of the supported Floquet--Bloch modes, providing a direct representation of the propagation characteristics of dispersive time-varying media, including spatiotemporal gaps and modulation-induced amplification, as schematically illustrated in Figures 1b--d. The eigenvectors define a modal basis in the resulting finite-dimensional complex vector space and describe the electromagnetic states supported by the dispersive time-varying medium. The corresponding system operator is given by
\begin{equation}
    \textbf{M}
    =
    \begin{bmatrix}
    \textbf{M}_1 & \textbf{0} \\
    \textbf{0} & \textbf{M}_2
    \end{bmatrix},
\end{equation}
where $\textbf{0}$ denotes the null matrix, and $\mathbf{M}_1$ and $\mathbf{M}_2$ are defined as
\begin{subequations}
    \begin{align}
    \textbf{M}_1
    &=
       \frac{1}{c^{2}}\,
    \boldsymbol{\omega}^{2}\,
    \left(\textbf{I}
    +
    \boldsymbol{\chi}\right)
    -
    \textbf{K}\,
    \left(\textbf{I}
    +
    \boldsymbol{\chi}\right)^{-1}\,
    \textbf{K}\,
    \left(\textbf{I}
    +
    \boldsymbol{\chi}\right),\\
    \textbf{M}_2
    &=
    \frac{1}{c^{2}}\,
    \boldsymbol{\omega}^2\,
    \left(\mathbf{I}
    +
    \boldsymbol{\chi}\right)\,
    -
    \mathbf{K}^{2}.
    \end{align}
\end{subequations}
Here, $\displaystyle c=1/\sqrt{\mu_0\varepsilon_0}$ is the speed of light in vacuum, $\mathbf{I}$ is the identity matrix, and $\boldsymbol{\omega}=\operatorname{diag}\left[\omega+n\Omega\right]$ and $\textbf{K}=\operatorname{diag}\left[k_x+mK\right]$ are diagonal operators in the spatiotemporal harmonic basis. The matrix $\boldsymbol{\chi}$ is the frequency-dependent susceptibility convolution operator, whose elements are defined as
\begin{equation}
    \boldsymbol{\chi}_{\nu\nu'}
    =
    \hat{\chi}_{n-n',\,m-m'}\!\left(\omega_{n'}\right),
\end{equation}
where $\nu$ and $\nu'$ label the harmonics associated with $(n,m)$ and $(n',m')$, respectively. The Floquet--Bloch wavenumbers follow from the spectrum of $\mathbf{M}$ as $\beta_q=\sqrt{\lambda_q}$, where $\lambda_q$ is the $q$-th eigenvalue of $\mathbf{M}$. Under the adopted Fourier convention, the appropriate Riemann sheet is chosen such that $\Im(\beta_q)\ge 0$ to enforce the radiation condition. The corresponding eigenvectors $\boldsymbol{\Psi}_e$ provide a complete modal basis for expanding the electric field within the truncated spatiotemporal harmonic space. The associated magnetic eigenvectors are then determined from Maxwell's equations as
\begin{equation}
    \boldsymbol{\Psi}_h
    =
    \begin{bmatrix}
    \mathbf{0}
    &
    -\varepsilon_0\boldsymbol{\omega}(\mathbf{I}+\boldsymbol{\chi})
    +\mu_0^{-1}\mathbf{K}\boldsymbol{\omega}^{-1}\mathbf{K}
    \\
    \varepsilon_0\boldsymbol{\omega}(\mathbf{I}+\boldsymbol{\chi})
    &
    \mathbf{0}
    \end{bmatrix}
    \boldsymbol{\Psi}_e
    \boldsymbol{\beta}^{-1},
\end{equation}
where $\displaystyle{\boldsymbol{\Psi}_h = \left[\textbf{H}^{(x)},\,\textbf{H}^{(y)}\right]^{\mathrm{T}}}$ collects the magnetic-field eigenvectors. Together, $\boldsymbol{\Psi}_e$ and $\boldsymbol{\Psi}_h$ define the electromagnetic eigenbasis of the dispersive time-varying medium and provide the foundation for extending the formulation to layered metasurfaces operating under fast spatiotemporal modulation (see Supporting Information, Section 2 for the detailed derivation of the eigenvalue equation).
\subsection{Floquet--Bloch Scattering Theory for Spatiotemporally Modulated Metasurfaces}
We consider a metasurface composed of a finite stack of spatiotemporally modulated layers arranged along the propagation direction $z$. The scattering construction adopted here is conceptually related to the RCWA framework widely used for passive spatially periodic metasurfaces, but extends this harmonic modal picture to causal dispersive systems with nonadiabatic temporal modulation \cite{inampudi2018rigorous}. As illustrated in Figure 1a, each layer supports its own Floquet--Bloch modal spectrum, determined by the electric susceptibility profile $\hat{\chi}_{e}^{(l)}$ and thickness $t_l$. Within each layer, the Floquet--Bloch eigenmodes form the natural basis for the electromagnetic fields. Therefore, the spatiotemporal harmonic components are represented as a superposition of forward and backward modal waves \cite{stratton2007electromagnetic},
\begin{subequations}
    \begin{align}
    \mathbf{E}_{nm}^{(l)}(z)
    &=
    \sum_{q}
    \left[
    a_{q}^{(l)}\,\boldsymbol{\Psi}_{e,nm}^{(l)}(q)\,e^{i\beta_{q}^{(l)}\left(z - z_l\right)}
    +
    b_{q}^{(l)}\,\boldsymbol{\Psi}_{e,nm}^{(l)}(q)\,e^{-i\beta_{q}^{(l)}\left(z - z_{l+1}\right)}
    \right], \quad z_l < z < z_{l+1},\\
    \mathbf{H}_{nm}^{(l)}(z)
    &=
    \sum_{q}
    \left[
    a_{q}^{(l)}\,\boldsymbol{\Psi}_{h,nm}^{(l)}(q)\,e^{i\beta_{q}^{(l)}\left(z - z_l\right)}
    -
    b_{q}^{(l)}\,\boldsymbol{\Psi}_{h,nm}^{(l)}(q)\,e^{-i\beta_{q}^{(l)}\left(z - z_{l+1}\right)}
    \right], \quad z_l < z < z_{l+1}.
    \end{align}
\end{subequations}
Here, $q$ indexes the Floquet--Bloch eigenmodes of layer $l$, $\boldsymbol{\Psi}_{e,nm}^{(l)}(q)$ and $\boldsymbol{\Psi}_{h,nm}^{(l)}(q)$ are the electric and magnetic harmonic components of the corresponding modal eigenvectors, $\beta_{q}^{(l)}$ is the associated propagation constant, and $a_{q}^{(l)}$ and $b_{q}^{(l)}$ denote the forward and backward modal amplitudes. The modal structure of each layer admits the compact state-vector representation
\begin{equation}
    \begin{bmatrix}
    \mathbf{E}^{(l)}(z) \\
    \mathbf{H}^{(l)}(z)
    \end{bmatrix}
    =
    \begin{bmatrix}
    \boldsymbol{\Psi}_{e}^{(l)} & \boldsymbol{\Psi}_{e}^{(l)} \\
    \boldsymbol{\Psi}_{h}^{(l)} & -\boldsymbol{\Psi}_{h}^{(l)}
    \end{bmatrix}
    \begin{bmatrix}
    e^{i\boldsymbol{\beta}^{(l)}\left(z - z_l\right)}\mathbf{a}^{(l)} \\
    e^{-i\boldsymbol{\beta}^{(l)}\left(z - z_{l +1}\right)}\mathbf{b}^{(l)}
    \end{bmatrix},
    \quad z_l < z < z_{l+1}.
\end{equation}
where the exponential factors are understood as diagonal propagation matrices acting on the modal amplitude vectors.\\
The continuity of tangential electromagnetic fields imposes a linear constraint between the incoming and outgoing modal amplitudes at each interface, which defines the interface scattering matrix as \cite{yeh1990optical}
\begin{equation}
\begin{bmatrix}
\mathbf{b}^{(l)} \\
\mathbf{a}^{(l+1)}
\end{bmatrix}
=
\mathbf{S}^{(l)}
\begin{bmatrix}
\boldsymbol{\Phi}_{l}\mathbf{a}^{(l)} \\
\boldsymbol{\Phi}_{l+1}\mathbf{b}^{(l+1)}
\end{bmatrix}
=
\begin{bmatrix}
\mathbf{S}_{11}^{(l)} & \mathbf{S}_{12}^{(l)} \\
\mathbf{S}_{21}^{(l)} & \mathbf{S}_{22}^{(l)}
\end{bmatrix}
\begin{bmatrix}
\boldsymbol{\Phi}_{l}\mathbf{a}^{(l)} \\
\boldsymbol{\Phi}_{l+1}\mathbf{b}^{(l+1)}
\end{bmatrix},
\end{equation}
in which $\mathbf{S}^{(l)}$ is the scattering matrix associated with the interface between layers $l$ and $l+1$, while
$\boldsymbol{\Phi}_{l}=\exp\!\left(i\boldsymbol{\beta}^{(l)}t_l\right)$
is the diagonal propagation matrix accounting for modal phase accumulation across layer $l$. The diagonal sub-blocks $\mathbf{S}_{11}^{(l)}$ and $\mathbf{S}_{22}^{(l)}$ represent reflection from layers $l$ and $l+1$, respectively, while the off-diagonal sub-blocks $\mathbf{S}_{21}^{(l)}$ and $\mathbf{S}_{12}^{(l)}$ represent transmission in the forward and backward directions. The interface scattering matrix is fully determined by the modal matching matrices
\begin{equation}
\begin{aligned}
\mathbf{A}_l &= \frac{1}{2}\left[
\left(\boldsymbol{\Psi}_e^{(l+1)}\right)^{-1}\boldsymbol{\Psi}_e^{(l)}
+
\left(\boldsymbol{\Psi}_h^{(l+1)}\right)^{-1}\boldsymbol{\Psi}_h^{(l)}
\right], \\
\mathbf{B}_l &= \frac{1}{2}\left[
\left(\boldsymbol{\Psi}_e^{(l+1)}\right)^{-1}\boldsymbol{\Psi}_e^{(l)}
-
\left(\boldsymbol{\Psi}_h^{(l+1)}\right)^{-1}\boldsymbol{\Psi}_h^{(l)}
\right],
\end{aligned}
\end{equation}
which respectively describe the symmetric and antisymmetric matching of the electric and magnetic modal bases across the interface (the resulting expressions for the sub-blocks of $\mathbf{S}^{(l)}$ are provided in Supporting Information, Section 3). The collective response of the metasurface stack is obtained by cascading the interface scattering matrices through the Redheffer star product,
\begin{equation}
\mathbf{S}
=
\mathbf{S}^{(0)} \star
\boldsymbol{\Phi}^{(1)} \star \mathbf{S}^{(1)} \star
\cdots \star
\boldsymbol{\Phi}^{(L)} \star \mathbf{S}^{(L)},
\end{equation}
where $\mathbf{S}$ is the total scattering matrix of the metasurface stack, and $\boldsymbol{\Phi}^{(l)}$ denotes the scattering matrix associated with propagation through layer $l$, defined as
\begin{equation}
\boldsymbol{\Phi}^{(l)}
=
\begin{bmatrix}
\mathbf{0} & \boldsymbol{\Phi}_l \\
\boldsymbol{\Phi}_l & \mathbf{0}
\end{bmatrix}.
\end{equation}
The matrices $\mathbf{S}^{(0)}$ and $\mathbf{S}^{(L)}$ describe the interfaces between the finite metasurface stack and the ambient regions on the incident and transmitted sides, respectively. Here, $\star$ denotes the Redheffer star product, which combines two cascaded scattering matrices while accounting for multiple internal reflections (the explicit block-matrix definition is provided in Supporting Information, Section 3).\\
The resulting scattering matrix $\mathbf{S}$ encapsulates the full spatiotemporal electromagnetic response of the metasurface, inherently accounting for dispersive harmonic coupling, and enables both the analysis of its interaction with external excitations and the identification of its intrinsic Floquet--Bloch modes. This makes the formulation particularly useful for PTCs, where the modal spectrum governs gap formation and parametric amplification.
\subsection{Floquet--Bloch Poynting Theorem and Observable Scattering Characteristics}
The scattering matrix formalism provides a rigorous framework for relating the Floquet--Bloch modal amplitudes across the metasurface. In order to connect this modal response to measurable scattering quantities, we introduce the time-averaged Poynting vector as \cite{stratton2007electromagnetic}
\begin{equation}
\langle \mathbf{S}(\mathbf{r}) \rangle
=
\frac{1}{T}
\int_{T}^{}
\mathbf{E}(\mathbf r,t)\times \mathbf{H}(\mathbf r,t)\, dt,
\end{equation}
where $\mathbf{E}(\mathbf r,t)$ and $\mathbf{H}(\mathbf r,t)$ denote the real time-domain fields. The resulting quantity defines the electromagnetic power flux and serves as the basis for evaluating the reflected and transmitted powers from the metasurface. Within the Floquet--Bloch representation, the orthogonality of the spatiotemporal harmonic functions allows the longitudinal power flux to be decomposed into contributions from individual harmonics,
\begin{equation}
P(z)
=
\frac{1}{2}
\sum_{n,m}
\Re
\left\{
\mathbf{E}_{nm}(z)
\times
\mathbf{H}_{nm}^*(z)
\right\}\cdot\hat{a}_z.
\end{equation}
\\
Using the modal expansion given in Equation (12), this harmonic power flux admits a compact quadratic form in terms of the forward and backward modal amplitudes,
\begin{equation}
P(z)
=
\frac{1}{2}
\Re
\left\{
\begin{bmatrix}
\boldsymbol{\Phi}_{+}\mathbf{a} \\
\boldsymbol{\Phi}_{-}\mathbf{b}
\end{bmatrix}^{\dagger}
\mathbf{Q}
\begin{bmatrix}
\boldsymbol{\Phi}_{+}\mathbf{a} \\
\boldsymbol{\Phi}_{-}\mathbf{b}
\end{bmatrix}
\right\},
\end{equation}
where $\boldsymbol{\Phi}_{\pm}(z)=\exp\left({\pm}i\boldsymbol{\beta}z\right)$ describe the longitudinal phase evolution of the forward and backward modes. The matrix $\mathbf{Q}$ defines the modal flux metric and is determined by the electric and magnetic Floquet--Bloch eigenvectors as
\begin{equation}
\mathbf{Q}
=
\begin{bmatrix}
\boldsymbol{\Psi}_{e} & \boldsymbol{\Psi}_{e}
\end{bmatrix}^{\dagger}
\begin{bmatrix}
\mathbf{0} & \mathbf{I} \\
-\mathbf{I} & \mathbf{0}
\end{bmatrix}
\begin{bmatrix}
\boldsymbol{\Psi}_{h} & -\boldsymbol{\Psi}_{h}
\end{bmatrix}.
\end{equation}
This quadratic form defines the flux metric of the dispersive Floquet--Bloch modal space and provides a direct means of calculating the longitudinal power flow from the modal amplitudes. In this representation, the electromagnetic power is obtained through the bilinear pairing between the electric and magnetic eigenvectors, while the modal amplitudes specify how the coupled spatiotemporal channels are populated during propagation.
\\
In particular, we assume the reflection and transmission regions to be homogeneous and lossless, with relative permittivities $\varepsilon^{(-)}$ and $\varepsilon^{(+)}$, respectively. In these regions, the Floquet--Bloch harmonics reduce to analytically defined plane-wave channels, with longitudinal propagation constants
\begin{equation}
\beta_{nm}^{\pm}
=
\sqrt{
\varepsilon^{(\pm)}
\left(\frac{\omega_n}{c}\right)^2
-
k_{m}^{2}
}.
\end{equation}
The square-root branch is chosen consistently with the radiation condition, such that propagating outgoing channels carry power away from the metasurface. As a direct consequence of the orthogonality of the plane-wave channels in the homogeneous lossless regions, the longitudinal power flux decomposes into independent contributions from the Floquet--Bloch harmonics,
\begin{equation}
P^{\pm}
=
\frac{1}{2}
\sum_{n,m}
\Re
\left\{
W_{nm}^{\pm}
\right\}
\left(
|a_{nm}^{\pm}|^2 - |b_{nm}^{\pm}|^2
\right).
\end{equation}
Here, $P^{\pm}$ denote the net signed longitudinal power fluxes in the homogeneous reflection and transmission regions, respectively, while $W_{nm}^{\pm}$ is the modal admittance factor of the $(n,m)$-th Floquet--Bloch harmonic in the corresponding region (see Supporting Information, Section 4 for additional details). Together, these relations establish a self-consistent formulation of power flow in dispersive time-varying metasurfaces.
\begin{figure}
    \centering
    \includegraphics[scale = 1]{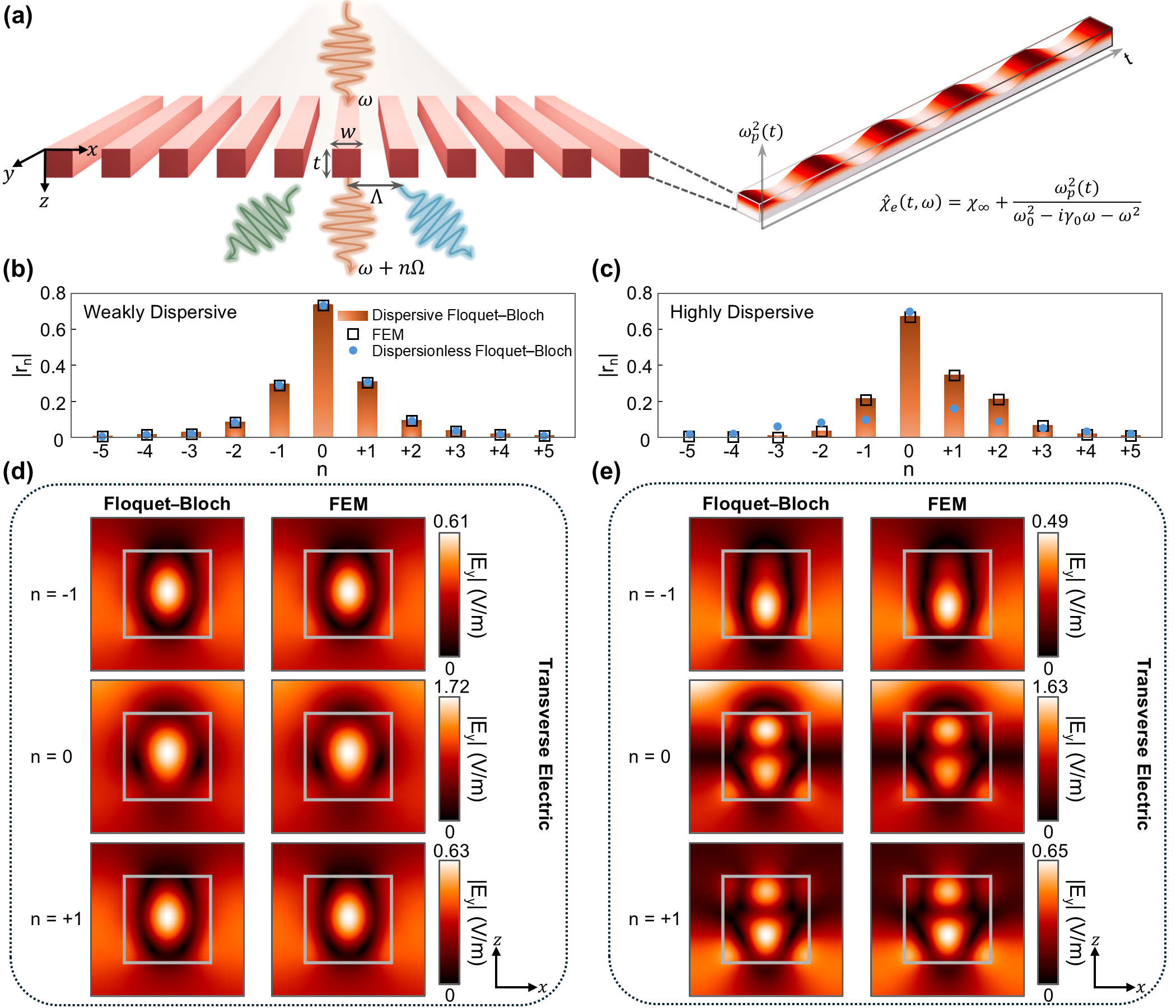}
    \caption{Numerical validation of the dispersive Floquet--Bloch theory for TE excitation. (a) Schematic of the free-standing dielectric grating used for validation. The grating material is dispersive and follows a Lorentzian response, while temporal modulation is introduced through a sinusoidally varying plasma frequency. (b,c) Harmonic reflection spectra for the weakly and highly dispersive scenarios, respectively, comparing the dispersive Floquet--Bloch formulation with FEM simulations and the dispersionless Floquet--Bloch model. (d,e) Spatial distributions of the electric field for the dominant harmonic orders in the weakly and highly dispersive scenarios, respectively, obtained using the Floquet--Bloch formulation and FEM simulations.}
    \label{fig.2}
\end{figure}
\section{Causal Dispersive Response and Numerical Validation}
Before embarking on the application of the developed Floquet--Bloch theory to address light--matter interaction with dispersive, time-varying metasurfaces, we first establish a causal material model that incorporates dispersion through a time-varying multipole Lorentz--Drude susceptibility \cite{mirmoosa2022dipole,ptitcyn2023floquet}. This model provides a physically consistent susceptibility kernel for the numerical examples and ensures that temporal modulation is introduced without violating the medium's causal response. To this end, we resort to the classical oscillator model for bound electrons, in which the polarization dynamics associated with the $j$-th Lorentz oscillator is governed by
\begin{equation}
\left[
\frac{\partial^2}{\partial t^2}
+
\gamma_j
\frac{\partial}{\partial t}
+
\omega_j^2
\right]
\mathbf{P}_j(\mathbf r,t)
=
\varepsilon_0 \omega_{p,j}^2(t)\mathbf{E}(\mathbf r,t),
\end{equation}
where $\omega_j$ and $\gamma_j$ are the resonance frequency and damping rate of the oscillator, respectively. The temporal modulation is introduced through the time-varying plasma frequency $\omega_{p,j}(t)$, allowing the oscillator strength, and therefore the dispersive susceptibility, to vary in time while respecting causality. After mathematical manipulation and by treating the delay variable via a Fourier transform, the dispersive time-varying susceptibility kernel is obtained as
\begin{equation}
\hat{\chi}_e(\mathbf r,t,\omega)
=
\chi_{\infty}
+
\sum_j
\frac{\omega_{p,j}^2(\mathbf r,t)}
{\omega_j^2-\omega^{\,2}-i\gamma_j\omega},
\end{equation}
where $\chi_{\infty}$ accounts for the background susceptibility associated with electronic transitions outside the spectral range of interest, while the summation describes the resonant Lorentz--Drude contributions. The conventional Drude response is recovered by setting the resonance frequency of the corresponding oscillator to zero, allowing the multipole Lorentz--Drude model to account for both bound and free charge carriers \cite{fox2010optical}. This causal material model therefore provides the basis for the numerical studies that follow, in which the developed Floquet--Bloch theory is applied to dispersive, time-varying metasurfaces and benchmarked against full-wave simulations.\\
In order to validate the proposed theory, we consider a free-standing dielectric grating with period $\Lambda$ along the $x$-direction, where the width and thickness of each ridge are set to $w=t=4\Lambda/7$, as shown in Figure 2a. The grating is composed of a dispersive time-varying medium characterized by a Lorentzian response with resonance frequency $\omega_0$, damping rate $\gamma_0$, static plasma frequency $\omega_{p,0}$, and background susceptibility $\chi_\infty=10.56$. Specifically, we introduce a sinusoidal temporal modulation to the plasma frequency as
\begin{equation}
\omega_p^2(t)
=
\omega_{p,0}^{2}
\left[
1+\delta\cos(\Omega t)
\right],
\end{equation}
where $\delta=0.8$ is the modulation depth and $\Omega$ is the angular modulation frequency. Under this sinusoidal temporal modulation profile, the oscillator strength contains only the static ($n=0$) and first-order ($n=\pm{1}$) temporal Fourier components, so the polarization at a given Floquet frequency is driven by the corresponding temporal harmonic and its two nearest neighbors,
\begin{equation}
\mathbf{P}_n(\mathbf r)
=
\varepsilon_0\chi(\omega_n)\mathbf{E}_n(\mathbf{r})
+
\varepsilon_0\frac{\delta}{2}\chi_L(\omega_n)
\left[
\mathbf{E}_{n-1}(\mathbf r)
+
\mathbf{E}_{n+1}(\mathbf r)
\right],
\end{equation}
in which $\chi(\omega_n)=\chi_\infty+\chi_L(\omega_n)$, with $\chi_L(\omega_n)$ standing for the static Lorentzian susceptibility evaluated at $\omega_n$. This nearest-neighbor selection rule in the Floquet ladder provides a direct basis for understanding the resulting harmonic spectra, while clearly revealing the role of material dispersion. Furthermore, in order to independently assess the consistency of the formulation, Equation (27) is implemented in COMSOL using the finite element method (FEM), and the resulting full-wave simulations are compared with the Floquet--Bloch calculations (details of the numerical implementation are provided in Supporting Information, Section 5).\\
\begin{figure}
    \centering
    \includegraphics[scale = 1]{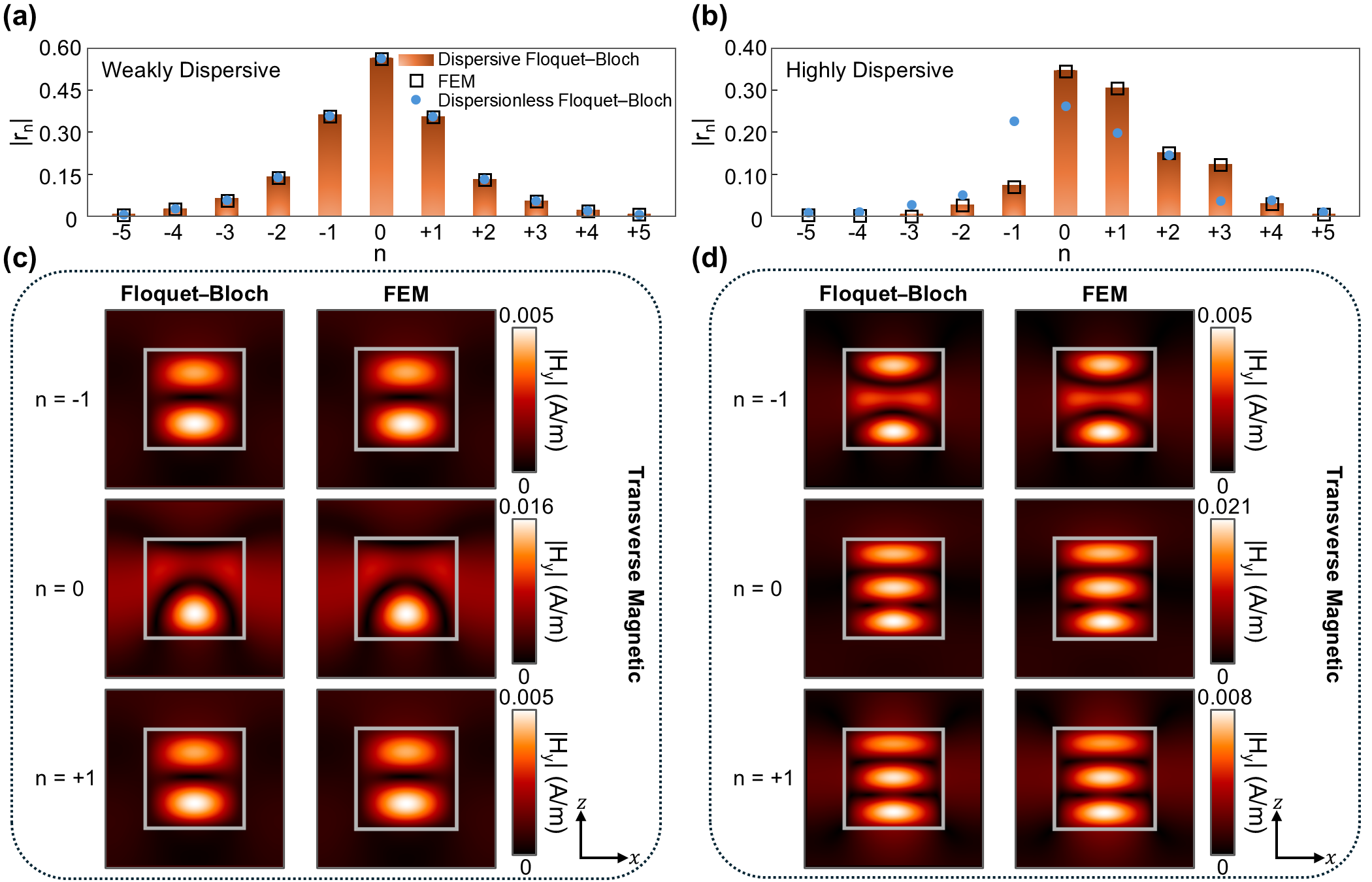}
    \caption{Numerical validation of the dispersive Floquet--Bloch theory for TM excitation. (a,b) Harmonic reflection spectra for the weakly and highly dispersive scenarios, respectively, comparing the dispersive Floquet--Bloch formulation with FEM simulations and the dispersionless Floquet--Bloch model. (c,d) Magnetic-field distributions for the dominant harmonic orders in the weakly and highly dispersive scenarios, respectively, obtained using the Floquet--Bloch formulation and FEM simulations.}
    \label{fig.3}
\end{figure}
In order to clearly demonstrate the role of material dispersion, we consider weakly and highly dispersive Lorentzian responses for both transverse electric (TE, or $s$-polarized) and transverse magnetic (TM, or $p$-polarized) waves, associated with the field components $E_y$ and $H_y$, respectively. The weakly dispersive regime corresponds to excitation away from the Lorentz resonance and a smaller modulation frequency, such that the susceptibility varies slowly across the coupled Floquet harmonics, whereas the highly dispersive regime is chosen closer to resonance and with a larger modulation frequency, so that different harmonics experience distinct material responses. This distinction allows us to isolate the effect of dispersion on harmonic generation, rather than attributing the response solely to the modulation strength. Since the grating is one-dimensional and invariant along $y$, the TE and TM polarization states are decoupled and can be treated independently. The block-diagonal form of $\mathbf{M}$ in Equation (8) follows from this decoupling. Therefore, the validation is organized into four representative cases, corresponding to weakly and highly dispersive responses under TE and TM excitation. In addition to the full dispersive Floquet--Bloch formulation, we also compute the response using its dispersionless counterpart that neglects the frequency dependence of the susceptibility, providing a direct comparison that highlights the importance of material dispersion (see Supporting Information, Section 5, for details on the dispersionless calculations). In each case, the excitation frequency is chosen in the vicinity of a fundamental guided-mode resonance of the grating, where the enhanced light–matter interaction amplifies the sensitivity of the generated Floquet harmonics to the underlying dispersive material response.\\
Figure 2 summarizes the validation results for TE excitation of the free-standing grating, where the structure is illuminated by a normally incident plane wave propagating along the $z$-direction with the electric field polarized along $y$. The incident wavelength is chosen as $\lambda=13\Lambda/7$, with the corresponding angular frequency denoted by $\omega$. In the weakly dispersive case, the material parameters are set to $\omega_0=\omega/3$, $\gamma_0=\omega_0/12$, and $\omega_{p,0}=4\omega_0$, while the modulation angular frequency is chosen as $\Omega=0.01\omega_0$, corresponding to a relatively slow modulation compared with the excitation frequency. As shown in Figure 2b, the harmonic reflection spectrum is dominated by the fundamental component and the first-order sidebands, associated with $n=0$ and $n=\pm1$, which is also suggested by the nearest-neighbor selection rule imposed by the sinusoidal modulation. Moreover, since the modulation frequency is small and the Lorentzian susceptibility varies weakly across the generated harmonics, the sidebands with opposite signs remain nearly symmetric, and the dispersionless Floquet--Bloch model provides a reasonable approximation to the full dispersive response. In contrast, the highly dispersive scenario, which is characterized by $\omega_0=1.1\omega$, $\gamma_0=0.01\omega_0$, $\omega_{p,0}=1.1\omega_0$, and $\Omega=0.1\omega_0$, exhibits a significantly different behavior. As can be seen in Figure 2c, the full dispersive Floquet--Bloch theory predicts an asymmetric distribution of the generated harmonics, which is in excellent agreement with FEM simulations, due to the fact that the Floquet harmonics associated with $n=+1$ and $n=-1$ sample different regions of the Lorentzian material response, with the larger modulation frequency further enhancing their spectral separation compared with the weakly dispersive scenario. By omitting this frequency dependence, the dispersionless formulation instead gives an almost symmetric harmonic distribution, illustrating the necessity of accounting for dispersion in the susceptibility kernel. In order to further assess the accuracy of the proposed formalism, the field maps in Figures 2d and 2e compare the spatial distributions of the electric field for the dominant harmonic orders $n=0$ and $n=\pm 1$ in the weakly and highly dispersive scenarios. In the weakly dispersive case, the harmonic fields retain similar modal profiles associated with the same grating resonance, whereas in the highly dispersive scenario the larger modulation frequency and stronger material dispersion lead to visibly distinct field distributions across different harmonics. The agreement between the Floquet--Bloch fields and FEM simulations confirms that the proposed formulation captures both the harmonic amplitudes and the spatial structure of the generated fields.\\
We next examine the same validation procedure for TM excitation, where the incident plane wave is characterized by a magnetic field polarized along $y$. The Lorentzian parameters defining the weakly and highly dispersive regimes are kept the same as in the TE case, while the excitation wavelength is chosen as $\lambda=11\Lambda/7$, with the corresponding angular frequency selected in the vicinity of the fundamental TM guided-mode resonance of the grating. The weakly dispersive response in Figure 3a again shows a harmonic reflection spectrum dominated by the fundamental order and the first-order sidebands, corresponding to $n=0$ and $n=\pm1$, respectively. Similar to the TE case, the relatively slow modulation and weak variation of the susceptibility across the generated harmonics lead to an almost symmetric sideband distribution, for which the dispersionless calculation remains close to the full dispersive Floquet--Bloch result. In the highly dispersive case shown in Figure 3b, however, the full dispersive theory reproduces the asymmetric harmonic distribution obtained from FEM, whereas the dispersionless calculation is unable to capture the relative strength of the generated sidebands as it neglects the frequency dependence of the Lorentzian response. The magnetic-field distributions in Figures 3c and 3d further confirm this behavior by comparing the dominant harmonic orders in the weakly and highly dispersive scenarios. While the harmonic fields retain similar modal profiles in the weakly dispersive regime, the larger modulation frequency and stronger material dispersion in the highly dispersive regime lead to harmonic-dependent field profiles. Not only does the agreement between the Floquet--Bloch and FEM results verify the harmonic reflection spectra, but it also confirms that the proposed dispersive formulation accurately captures the near-field structure of the generated TM harmonics.\\
In obtaining the above results using the Floquet--Bloch formulation, the temporal and spatial harmonic expansions were truncated at $N=10$ and $M=50$, respectively. This truncation was sufficient to ensure convergence of the harmonic spectra and near-field distributions (see Supporting Information, Section 5 for details of the convergence analysis). Notably, even for the simple sinusoidal modulation considered here, the Floquet--Bloch formulation provides a computationally efficient route to the steady-state harmonic response, whereas more general full-wave time-domain simulations would require resolving the temporal dynamics over many modulation periods and can therefore become highly time-consuming. Having verified the accuracy of the formulation against full-wave simulations, we now use the resulting scattering matrix to address complementary aspects of dispersive time-varying metasurfaces, ranging from their interaction with external excitations to the intrinsic modal structure encoded in the singularities of the scattering response.
\begin{figure}
    \centering
    \includegraphics[scale = 1]{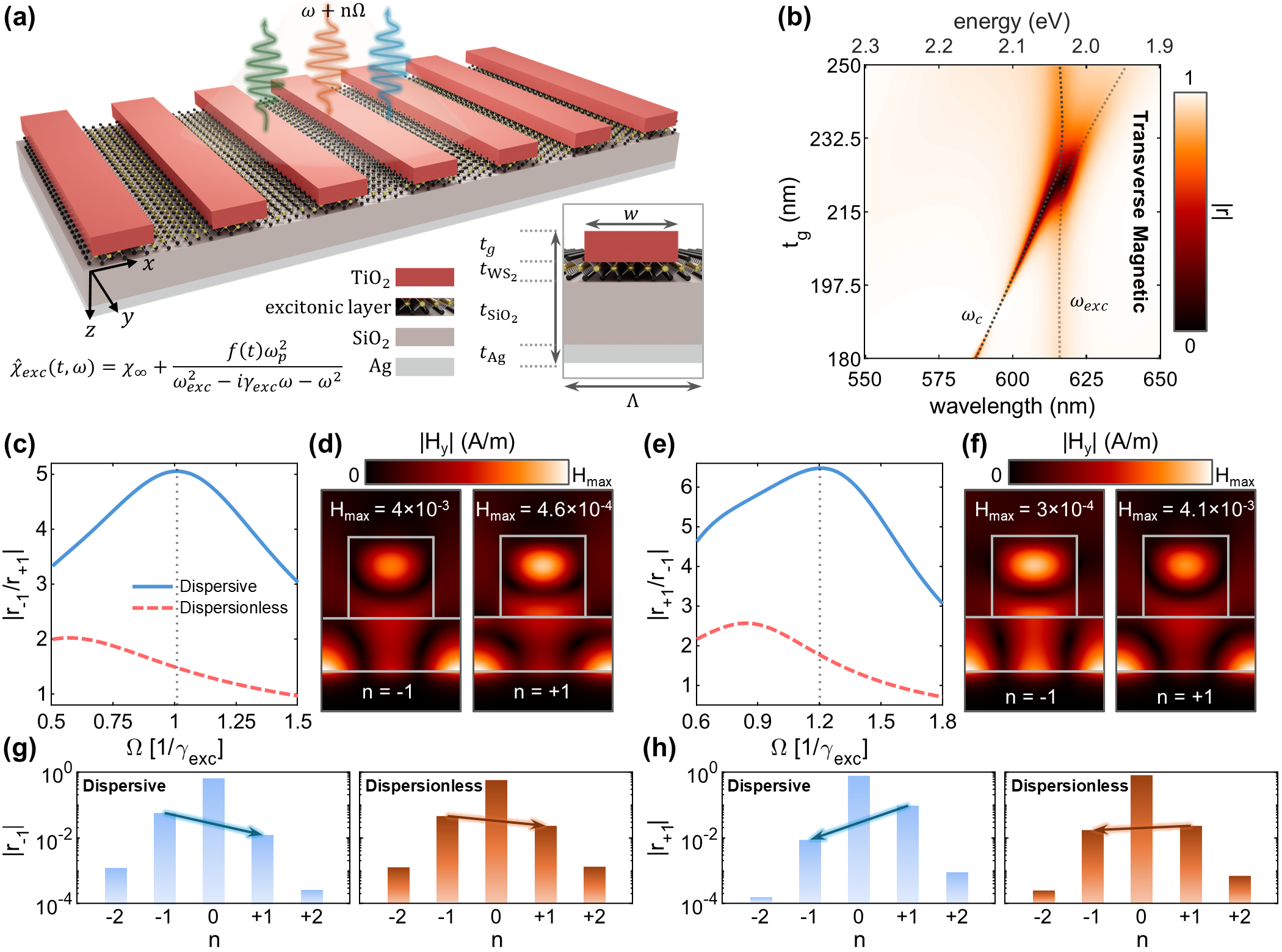}
    \caption{Asymmetric Floquet harmonic generation in a time-modulated excitonic metasurface based on a monolayer excitonic material. (a) Schematic of the metasurface consisting of a TiO$_2$ grating placed on a monolayer WS$_2$, separated from an Ag mirror by a SiO$_2$ spacer. The structure is excited by a normally incident TM-polarized plane wave, and the oscillator strength of the excitonic susceptibility is temporally modulated as $f(t)=1+\delta\cos(\Omega t)$. (b) Static reflection response of the unmodulated metasurface as a function of wavelength and TiO$_2$ grating thickness, showing the coupling between the TM cavity resonance and the excitonic resonance marked by the vertical dashed line. (c,e) Harmonic asymmetry ratios for the blue- and red-detuned scenarios, respectively, compared with the dispersionless model. (d,f) Magnetic-field distributions of the $n=-1$ and $n=+1$ harmonics for the blue- and red-detuned scenarios, showing selective enhancement of the down- and up-converted harmonics, respectively. (g,h) Harmonic reflection spectra for the blue- and red-detuned scenarios, respectively, calculated using the dispersive and dispersionless models.}
    \label{fig.4}
\end{figure}
\section{Results and Discussion}
In this section, after introducing the causal dispersive model and validating the developed theory against full-wave finite-element simulations in weakly and highly dispersive regimes, we apply the theory to a time-modulated excitonic metasurface in order to emphasize the impact of dispersion in nonadiabatic time-varying systems and demonstrate asymmetric Floquet harmonic generation driven by the strongly dispersive excitonic response. Finally, to demonstrate the capability of the theory to resolve the underlying modal structure under fast modulation, we study a time-varying metasurface as a PTC, using the theory to show how its momentum bandgap emerges from the hybridization of counter-propagating Floquet--Bloch modes. Together, these examples reveal various physical aspects of the theory, ranging from dispersive harmonic generation to modal engineering of band structure.
\subsection{Asymmetric Floquet Harmonic Generation in a Time-Modulated Excitonic Metasurface}
Having validated the theoretical formalism, herein we study, as an illustrative example, asymmetric Floquet harmonic generation enabled by the highly dispersive response of a time-varying excitonic metasurface. Such excitonic behavior is particularly well suited for demonstrating the capability of the developed theory, as the frequency-dependent light--matter interaction can vary strongly across different Floquet channels under fast modulation conditions. As mentioned earlier, nonadiabatic time modulation in resonant dispersive media brings about a spectrum of sidebands whose distribution is imposed not only by the temporal modulation profile but also by the frequency-dependent light--matter interaction experienced by each Floquet channel. Therefore, engineering the resonant optical response of the metasurface provides a direct means of controlling the relative strength of the generated Floquet harmonics and tailoring the spectral content of the scattered field. More specifically, the excitonic response of two-dimensional semiconductors can be dynamically tailored through various mechanisms such as the optical Stark effect, electrostatic doping, and strain-induced modulation, enabling dynamic control over their resonant susceptibility in photonic structures \cite{sie2017valley,cunningham2019resonant,chakraborty2018control,he2016strain}. In this context, monolayer WS$_2$ serves as a representative example of an excitonic material, as its prominent A-exciton resonance in the visible range provides a highly dispersive optical response that can be used to examine dispersion-mediated Floquet harmonic generation. Here, we demonstrate that a simple sinusoidal temporal modulation, when applied to an excitonic layer integrated with a resonant metasurface, is sufficient to enable dispersion-mediated control of Floquet harmonic generation, where the excitonic resonance lifts the symmetry between up- and down-converted scattering channels and selectively favors one generated harmonic over its counterpart. We further emphasize the importance of the dispersive Floquet--Bloch theory by showing that a dispersionless model fails to capture this spectral imbalance and instead predicts nearly symmetric up- and down-converted Floquet harmonics.\\
Figure 4a illustrates the schematic of the metasurface, where a TiO$_2$ grating is placed on top of a monolayer excitonic layer, with a SiO$_2$ spacer separating it from an Ag mirror. The structure is excited by a normally incident TM-polarized plane wave with angular frequency $\omega$, characterized by a magnetic field polarized along the $y$-direction. For simplicity, TiO$_2$ and SiO$_2$ are modeled as dispersionless dielectrics with refractive indices of 2.4 and 1.46, respectively. The excitonic layer is taken to have a thickness $t_{\mathrm{WS}_2}=0.7~\mathrm{nm}$, representative of monolayer WS$_2$, and its susceptibility is approximated by a single Lorentzian oscillator, as the A-exciton dominates the material response over the spectral range of interest \cite{cong2018optical,bianchi2024engineering}. The Lorentzian model is characterized by an exciton energy $\hbar\omega_{\mathrm{exc}}=2.01~\mathrm{eV}$, linewidth $\hbar\gamma_{\mathrm{exc}}=30~\mathrm{meV}$, and resonance strength determined by $\hbar\omega_p=1.38~\mathrm{eV}$, while a nonresonant background susceptibility $\chi_\infty=17$ accounts for optical transitions outside the spectral range considered here \cite{li2014measurement,weber2023intrinsic}. Time modulation in the metasurface is introduced through the oscillator strength of the monolayer excitonic susceptibility, defined as
\begin{equation}
f(t)=1+\delta\cos(\Omega t),
\end{equation}
where \(\Omega\) is the angular modulation frequency. The modulation depth is chosen as \(\delta=0.6\) to clearly resolve the modulation-induced harmonic asymmetry while keeping the oscillator strength positive throughout the modulation cycle. This choice is intended as a theoretical modulation profile for highlighting the role of excitonic dispersion in Floquet harmonic generation, rather than as a specification of a particular experimental modulation scheme. To translate the time-dependent excitonic response into a pronounced Floquet scattering response, the excitonic layer is embedded in a resonant photonic environment, where enhanced field confinement increases the overlap between the optical mode and the atomically thin material. Specifically, the TiO$_2$ grating acts as a photonic cavity that couples to the exciton and enhances the dispersive optical response of the metasurface, thereby strengthening the interaction between the incident field and the generated Floquet harmonics. By setting $\delta=0$, we first characterize this cavity--exciton coupling in the static structure and use the resulting reflection response as a spectral reference for the time-modulated results discussed later in this section.\\
In order to bring a fundamental guided mode of the TiO$_2$ grating to the vicinity of the excitonic resonance, we choose a grating structure with spatial period $\Lambda=400~\mathrm{nm}$ and groove width $w_g=280~\mathrm{nm}$, which allows the resonant wavelength to be tuned through the grating thickness $t_g$. The Ag layer thickness is chosen as $t_{\mathrm{Ag}}=120~\mathrm{nm}$, exceeding the optical skin depth of silver in the spectral range of interest and effectively suppressing transmission through the back reflector, while the SiO$_2$ spacer thickness is set to $t_{\mathrm{SiO}_2}=150~\mathrm{nm}$ to control the coupling between the guided mode and the Ag mirror. We then utilize the grating thickness to spectrally couple a TM-guided mode of the grating to the excitonic resonance of the monolayer. The TM polarization state is particularly beneficial in this configuration, as the discontinuity of the normal component of the electric field at the grating boundaries enhances field localization near the excitonic layer, thereby strengthening its interaction with the resonant optical response. As illustrated in Figure 4b, the static reflection response demonstrates the trajectory of the TM cavity resonance, with frequency $\omega_c$, as the grating thickness is varied, while the excitonic resonance is indicated by the vertical dashed line. As the cavity mode approaches the excitonic resonance, the reflection is strongly suppressed, suggesting enhanced absorption within the metasurface. This enhanced absorption reflects stronger coupling between the photonic cavity mode and the excitonic response, giving rise to a highly dispersive light--matter interaction that is later used to bias the generated Floquet harmonics. Accordingly, we set the grating thickness to $t_g=223~\mathrm{nm}$, where the cavity resonance enhances the excitonic contribution to the metasurface response and provides a favorable operating point for selectively enhancing a desired Floquet harmonic while suppressing its counterpart.\\
To elucidate the mechanism underlying dispersion-mediated asymmetric harmonic generation, we consider two complementary detuning scenarios in which the excitation frequency is placed on opposite sides of the excitonic resonance. We refer to these cases as the blue- and red-detuned scenarios, corresponding respectively to excitation above and below the excitonic resonance, such that the first-order Floquet harmonics of opposite signs couple unequally to the metasurface resonance. The emergence of asymmetric Floquet harmonics can be explained by the scattering matrix of the time-modulated metasurface, in which the resonant structure encodes not only the cavity resonance coupled to the excitonic response, but also the Floquet replicas generated by temporal modulation. Therefore, we utilize the modulation frequency as a tuning knob to shift the Floquet replicas relative to the excitation frequency, allowing a selected resonant branch to overlap spectrally with the incident wave and thereby enhance the desired converted harmonic. To quantify the harmonic asymmetry, Figure 4c shows $|r_{-1}/r_{+1}|$ as a function of the modulation angular frequency for the blue-detuned scenario, where the metasurface is excited at $\omega=\omega_{\mathrm{exc}}+0.64\gamma_{\mathrm{exc}}$. As $\Omega$ is varied, the harmonic ratio exhibits a peak at $\Omega=1.01\gamma_{\mathrm{exc}}$, indicating that this modulation frequency places a Floquet replica of the excitonic resonance near the excitation frequency. Under this resonant condition, the down-converted harmonic is selectively enhanced, resulting in an approximately fivefold contrast between $|r_{-1}|$ and $|r_{+1}|$. Notably, although the dispersive Floquet--Bloch theory predicts a prominent asymmetry between the generated harmonics, the dispersionless model fails to capture this feature because it omits the frequency-dependent role of the excitonic resonance in the Floquet ladder. This asymmetry is further reflected in the magnetic fields shown in Figure 4d, where the field profiles associated with $n=-1$ and $n=+1$ exhibit similar spatial patterns, suggesting coupling to the same photonic cavity mode. However, the $n=-1$ harmonic is nearly an order of magnitude stronger than its $n=+1$ counterpart, confirming that the observed asymmetry originates from selective resonant enhancement through the excitonic response rather than from a change in the underlying modal profile.\\
We then turn to the red-detuned scenario, where the metasurface is excited below the excitonic resonance at $\omega=\omega_{\mathrm{exc}}-\gamma_{\mathrm{exc}}$. In this case, Figure 4e shows the harmonic ratio $|r_{+1}/r_{-1}|$ as a function of the modulation angular frequency, with a clear maximum at $\Omega=1.21\gamma_{\mathrm{exc}}$. Consistent with the previous case, this peak indicates spectral overlap between the relevant Floquet replica of the excitonic resonance and the excitation frequency, while now favoring the up-converted Floquet harmonic instead of the down-converted enhancement observed in the blue-detuned scenario. Therefore, the up-converted harmonic becomes dominant, yielding an asymmetric response in which $|r_{+1}|$ exceeds $|r_{-1}|$ by approximately a factor of seven. The corresponding magnetic-field distributions in Figure 4f further show that the first-order harmonics retain similar mode profiles, while the $n=+1$ field is nearly an order of magnitude stronger than the $n=-1$ field. Thus, moving the excitation from the blue- to the red-detuned side of the excitonic resonance reverses the dominant Floquet harmonic from the down-converted to the up-converted channel. Similar to the previous scenario, the dispersionless model is incapable of predicting the asymmetric Floquet response and instead results in a nearly balanced harmonic distribution. To further highlight this contrast, the harmonic reflection spectra obtained from the dispersive and dispersionless models are shown on a logarithmic scale in Figures 4g and 4h for the blue- and red-detuned scenarios, respectively. In both cases, the reflected spectrum is dominated by the fundamental component and the first-order harmonics, corresponding to $n=0$ and $n=\pm1$, respectively. The dispersive Floquet--Bloch theory reveals a pronounced asymmetry between the two first-order sidebands, where the enhanced harmonic is nearly one order of magnitude larger than its counterpart. In contrast, the dispersionless counterpart results in an almost symmetric harmonic distribution, as it neglects the frequency-dependent excitonic response that governs the unequal coupling of the generated Floquet orders.\\
These results suggest that the excitation detuning, combined with the dispersive excitonic response of the metasurface, not only controls the strength of Floquet harmonic generation but also determines whether the up- or down-converted channel is selectively enhanced. In particular, the observed asymmetry is obtained under a simple sinusoidal modulation profile, indicating that it originates from resonant material dispersion rather than from a complex temporal waveform. This application highlights the importance of incorporating dispersion when modeling resonant time-varying excitonic media, whose optical response can vary noticeably across the generated Floquet harmonics. More broadly, it demonstrates the relevance of the dispersive Floquet--Bloch scattering theory for analyzing photonic systems in which material dispersion and temporal modulation are inherently intertwined.
\begin{figure}
    \centering
    \includegraphics[scale = 1]{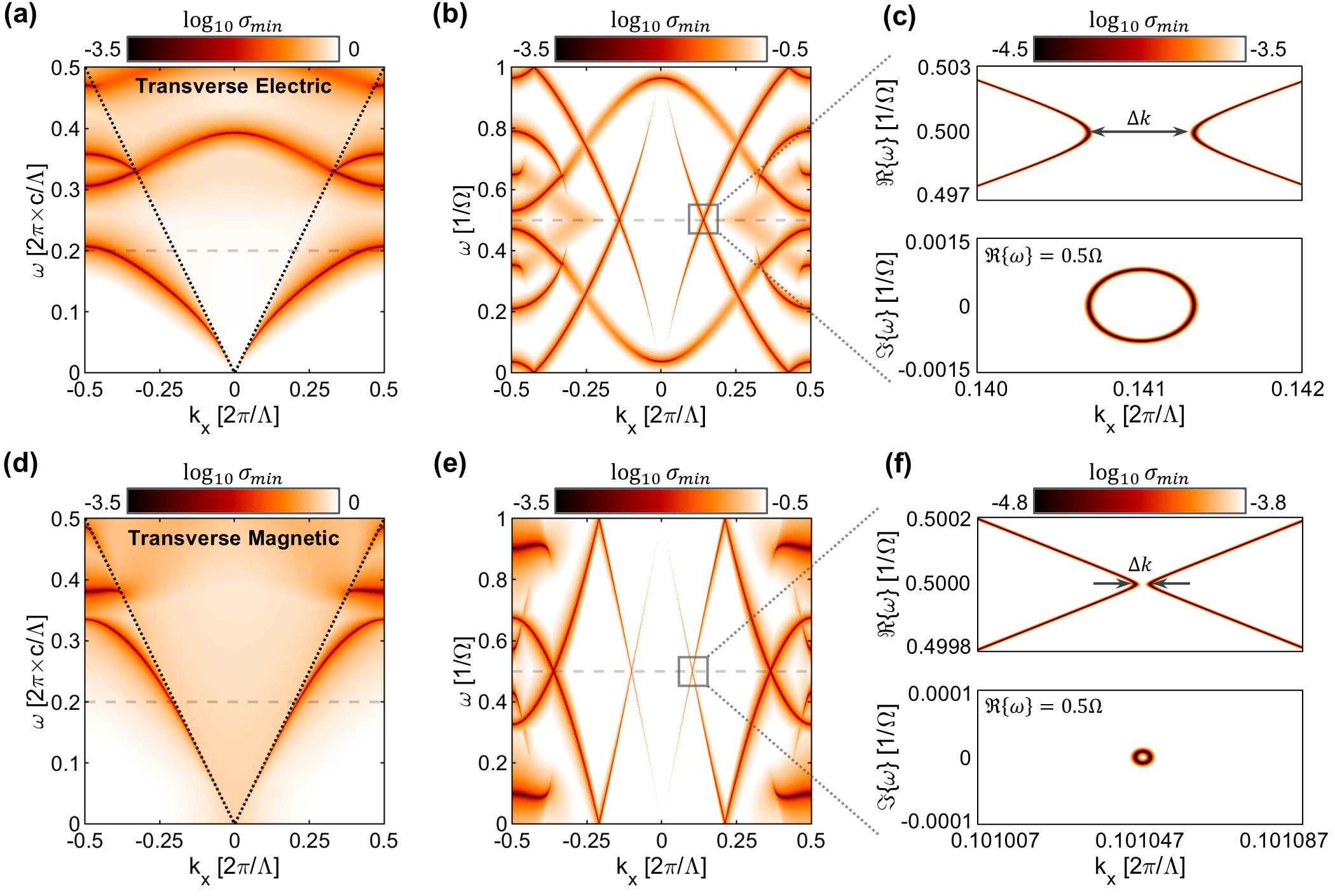}
    \caption{
    Band structures of a time-varying metasurface as a photonic time crystal, for TE and TM polarizations. (a,d) Static band structures of the metasurface for TE and TM modes, respectively, obtained at $\delta = 0$. The dashed gray lines indicate the selected modulation frequency, and the dashed black lines mark the boundaries of the folded light cone. (b,e) Band structures of the time-varying metasurface for TE and TM modes under temporal modulation with $\delta = 0.01$ and $\Omega = 0.2(2\pi c / \Lambda)$. (c,f) Magnified views of the momentum bandgaps for TE and TM near the center of the temporal Brillouin zone, together with the complex-frequency branches inside the gap.}
    \label{fig.5}
\end{figure}
\subsection{Photonic Time Crystals Enabled by Time-Varying Metasurfaces}
Photonic time crystals constitute the emerging temporal counterparts of conventional photonic crystals, replacing spatial periodicity with a periodic variation of optical properties in time. Therefore, the conserved and forbidden quantities are interchanged, and instead of opening frequency bandgaps at fixed wavevector, temporal periodicity brings about momentum bandgaps in which the eigenfrequencies become complex and electromagnetic modes either grow or decay in time \cite{asgari2024theory,boltasseva2024photonic}. Although this concept is commonly introduced for spatially uniform bulk media, its optical realization remains challenging, since pronounced momentum bandgaps require strong temporal coupling between shifted modal branches. In this sense, photonic resonances offer a viable path beyond this limitation, as they intensify this coupling and facilitate the opening of momentum bandgaps \cite{wang2025expanding,garg2025photonic,huang2026observation}. Time-varying metasurfaces are promising candidates for exploiting this resonant mechanism, as their spatial periodicity provides guided-mode branches that can be efficiently coupled to their Floquet counterparts through temporal modulation. This interaction between Bloch waves and temporal Floquet harmonics gives rise to Floquet--Bloch modes, which not only establish the metasurface as a PTC but also facilitate the formation of momentum bandgaps. To demonstrate this concept, we use the developed Floquet--Bloch theory to obtain the modal characteristics of a time-varying metasurface treated as a PTC. In this application, the quasinormal modes of the metasurface provide the underlying branches that are coupled by temporal modulation, leading to the opening of momentum bandgaps. The finite radiative lifetime of these modes is naturally retained in the modal description, allowing the band structure to capture both propagating and leaky branches of the metasurface. Our analysis suggests that a time-modulated metasurface can operate as a PTC, where spatially guided modes mediate temporal Floquet coupling and support momentum bandgaps with complex eigenfrequencies, enabling parametric amplification through temporal modulation.\\
As mentioned earlier, the Floquet--Bloch theory developed in this work can be utilized not only to describe the response of a time-varying metasurface under external illumination, but also to resolve its inherent modal characteristics by analyzing the corresponding scattering matrix. In this regard, we use the singularities of the scattering matrix to identify the modal branches of the system and access the spectrum of quasinormal modes. To quantify these singularities, we use the smallest singular value of the scattering matrix, which provides a stable metric for locating the modal branches. By tracking this quantity as a function of Bloch wavevector and frequency, the band structure of the time-varying metasurface can be extracted directly. This provides a natural route to treating the metasurface as a PTC and analyzing the formation of momentum gaps within the same framework used for scattering calculations. Here, we consider a free-standing dielectric grating similar to that shown in Figure 2a, with a spatial period $\Lambda$ along the $x$-axis, supporting guided-mode branches arising from the index contrast between the grating and the surrounding medium. Temporal modulation is introduced through the electric susceptibility of the dielectric grating, defined as
\begin{equation}
\chi_e(t)=\chi_0\left[1+\delta\cos(\Omega t)\right],
\end{equation}
allowing the Bloch branches of the static metasurface to couple to their Floquet counterparts and giving rise to momentum bandgaps. Although the developed theory can capture material dispersion, we consider a dispersionless material model in this application for clarity, since our primary objective is to examine the metasurface as a PTC. Within this setting, the modal spectrum directly reveals how temporal modulation reshapes the metasurface band structure and gives rise to momentum bandgaps.\\
We begin our analysis by setting $\delta=0$ to establish the dispersion of the Bloch modes supported by the static metasurface, which serves as the time-invariant photonic crystal underlying the subsequent Floquet coupling. For this reference structure, the grating width and thickness are chosen as $w=t=4\Lambda/7$, and the static electric susceptibility is set to $\chi_0=10.56$, which fixes the modal scale for the analysis. The modal dispersion is obtained by sweeping the Bloch wavevector over the first spatial Brillouin zone, $k_x\in[-K/2,K/2]$, and identifying the eigenfrequencies $\omega$ of the quasinormal-mode branches from the minima of the smallest singular value, $\sigma_{\min}$, of the system matrix. In the band diagrams shown here, we restrict the search to real eigenfrequencies, while the same procedure can be extended into the complex-frequency plane to recover the full spectrum of quasinormal modes. Figures 5a and 5d illustrate the band structures of the Bloch modes supported by the static photonic crystal for TE and TM polarization states, respectively. In the absence of temporal modulation, the band structures exhibit the characteristic Bloch dispersion of the metasurface, with the light cone folded into the first spatial Brillouin zone due to spatial periodicity. The supported modes appear as guided-mode branches within the folded light cone and remain symmetric with respect to the center of the spatial Brillouin zone. These branches constitute the modal structure that will be coupled through temporal modulation to form momentum gaps in the PTC band structure. Specifically, we focus on the first fundamental TE and TM guided-mode branches and choose the modulation frequency within this spectral range, as indicated by the dashed gray lines in the band diagrams, providing the reference point for examining how temporal modulation reshapes these branches and opens momentum bandgaps.\\
Next, we study the metasurface as a PTC by introducing temporal modulation as a small perturbation, characterized by $\delta=0.01$. For both TE and TM modes, the modulation angular frequency is chosen as $\Omega=0.2(2\pi c/\Lambda)$, placing it within the spectral range of the first fundamental guided-mode branch. In analogy with spatial periodicity, periodic temporal modulation folds the frequency into the first temporal Brillouin zone, $\omega\in[-\Omega/2,\Omega/2]$, so that the coupled modal branches can be analyzed within a single frequency period. The resulting band structures of the time-varying metasurface for TE and TM modes are illustrated in Figures 5b and 5e, respectively. As can be seen, temporal modulation couples the static Bloch branches to their Floquet counterparts and gives rise to Floquet--Bloch modes in the time-varying metasurface. Since the time-varying susceptibility $\chi_e(t)$ is real-valued and periodic in time, the resulting band structures exhibit mirror symmetry with respect to the center of the temporal Brillouin zone at $\omega=\Omega/2$. For both TE and TM modes, the interaction between the coupled modal branches produces avoided crossings at $\omega=\Omega/2$, corresponding to the center of the temporal Brillouin zone, as also occurs in conventional bulk PTCs. The location of these gaps follows from the temporal folding of the modal spectrum, so that the relevant coupling occurs where Floquet branches become degenerate within the first temporal Brillouin zone. These avoided crossings open momentum bandgaps, within which the eigenfrequencies become complex. Such complex-frequency branches are associated with amplification and attenuation of electromagnetic modes and provide a route to parametric amplification enabled by temporal modulation.\\
Figures 5c and 5f show magnified views of the momentum gaps for the TE and TM modes, together with the corresponding complex-frequency branches near the center of the temporal Brillouin zone. Inside the gap, $\Re(\omega)$ remains fixed at $\Re(\omega)=\Omega/2$, while the imaginary parts are recovered by extending the modal search into the complex-frequency plane. To quantify the gap size, we define $\Delta k$ as the separation between the two real-valued Bloch wavevectors that bound the momentum bandgap at $\Re(\omega)=\Omega/2$. As can be seen, for both TE and TM modes, the imaginary part of the eigenfrequency traces a closed contour resembling a semicircle, vanishing at the two real-valued Bloch wavevectors that bound the gap and reaching its maximum near the center of the momentum bandgap. This semicircular profile is expected for weakly modulated PTCs, where the imaginary branch follows from the perturbative coupling between degenerate Floquet replicas at $\Re(\omega)=\Omega/2$. Quantitatively, the relative momentum gap is $\Delta k/K\approx10^{-3}$ for the TE mode and $\Delta k/K\approx10^{-5}$ for the TM mode, differing by approximately two orders of magnitude. This contrast can be traced back to the different positions of the two modal branches relative to the light line at $\omega=\Omega/2$. As shown in Figure 5d, the TM branch is nearly linear around $\omega=\Omega/2$, indicating that the metasurface can be approximately spatially homogenized in this spectral region. In this limit, its response approaches that of a spatially uniform PTC, and temporal modulation opens only a narrow momentum gap. Figure 5a, on the other hand, shows that the TE branch has a more dispersive guided-mode profile in the same spectral region, indicating stronger modal interaction and producing a more pronounced momentum gap. Beyond the specific configuration considered here, this analysis underscores the role of Floquet--Bloch mode engineering in shaping momentum bandgaps and demonstrates the value of a self-contained framework that connects the scattering response of a time-varying metasurface to its modal band structure.\\
We have shown throughout this example that a time-varying metasurface can host Floquet--Bloch modes whose band structure carries the defining signatures of PTCs, including momentum bandgaps, complex eigenfrequency pairs, and the modal conditions associated with parametric amplification. By treating the scattering response and the modal band structure within a rigorous theory, the Floquet--Bloch formulation developed in this work provides direct access to the underlying physics of the PTC, allowing its defining features to be extracted from the eigenstructure of the time-varying metasurface rather than inferred solely from its externally driven response. Together, these results show that engineering the dispersion of the underlying Floquet--Bloch modes provides a viable pathway toward shaping PTC behavior in time-varying metasurfaces.
\section{Conclusion}
We have developed a comprehensive Floquet--Bloch theory to describe the response of nonadiabatic time-varying metasurfaces in the presence of material dispersion, with causality incorporated at the constitutive level through causal material response kernels consistent with the Kramers--Kronig relations. By projecting the electromagnetic fields onto a Floquet--Bloch basis, our proposed theory addresses the intertwined effects of spatial periodicity and temporal modulation, and the frequency-dependent coupling between spatiotemporal harmonics is taken into account via a dispersive susceptibility kernel. The modal picture is then extended to finite metasurface stacks through a scattering-matrix formalism, and a Floquet--Bloch Poynting theorem is derived to provide a consistent notion of power in dispersive, spatiotemporally modulated media.\\
We validated the results predicted by this theory against full-wave finite-element simulations in weakly and highly dispersive regimes, showing that a dispersionless formulation is incapable of capturing light--matter interaction phenomena that arise when different Floquet harmonics sample distinct regions of the material response. We further demonstrated asymmetric Floquet harmonic generation in a time-modulated excitonic metasurface, in which the strongly dispersive excitonic response, together with a guided-mode resonance, selectively enhances either the up- or down-converted Floquet sideband under a simple sinusoidal modulation profile. Next, we studied a time-varying metasurface as a PTC and obtained its inherent modal structure using the proposed Floquet--Bloch theory. Through this analysis, we established that the dispersion of the underlying quasinormal modes can be deliberately manipulated to control the resulting momentum bandgap, thereby providing a route toward engineering PTC behavior in resonant time-varying metasurfaces.\\
The introduced Floquet--Bloch theory is particularly beneficial, as it not only constitutes a self-contained framework to study the interaction of external excitations with dispersive, rapidly time-varying metasurfaces, but also provides insight into the underlying physics of their quasinormal modes and associated modal band structures. In analogy with RCWA for conventional passive metasurfaces, the formulation provides a modal scattering framework for layered structures, while going beyond this setting by incorporating fast temporal Floquet coupling, material dispersion, and access to the quasinormal-mode band structure of time-varying metasurfaces. This capability makes the framework especially valuable for analyzing resonant spatiotemporal systems in which scattering features, modal dispersion, and energy exchange are intrinsically connected, and provides a useful foundation for studying metasurface-based PTCs. More broadly, by unifying dispersion, modal physics, and scattering observables within a single Floquet--Bloch theory, this work provides a conceptual basis for spatiotemporal photonic platforms in which material response, structural design, and fast temporal modulation are inherently intertwined.

\newpage

\bibliography{References}

@article{galiffi2022photonics,
  title={Photonics of time-varying media},
  author={Galiffi, Emanuele and Tirole, Romain and Yin, Shixiong and Li, Huanan and Vezzoli, Stefano and Huidobro, Paloma A and Silveirinha, M{\'a}rio G and Sapienza, Riccardo and Al{\`u}, Andrea and Pendry, John B},
  journal={Advanced Photonics},
  volume={4},
  number={1},
  pages={014002--014002},
  year={2022},
  publisher={Society of Photo-Optical Instrumentation Engineers}
}

@article{patel2026photonic,
  title={Photonic Time Crystals and Time-Varying Electromagnetic Metamatter: A New Direction for Ultrafast Tunable Photonic and Microwave Materials and Devices},
  author={Patel, Ranjan Kumar and Ramanathan, Shriram and Jenkins, Ronald P and Carter, Michael J},
  journal={Advanced Science},
  pages={e19790},
  year={2026},
  publisher={Wiley Online Library}
}

@article{engheta2023four,
  title={Four-dimensional optics using time-varying metamaterials},
  author={Engheta, Nader},
  journal={Science},
  volume={379},
  number={6638},
  pages={1190--1191},
  year={2023},
  publisher={American Association for the Advancement of Science}
}

@article{asgari2024theory,
  title={Theory and applications of photonic time crystals: a tutorial},
  author={Asgari, Mohammad M and Garg, Puneet and Wang, Xuchen and Mirmoosa, Mohammad S and Rockstuhl, Carsten and Asadchy, Viktar},
  journal={Advances in optics and photonics},
  volume={16},
  number={4},
  pages={958--1063},
  year={2024},
  publisher={Optica Publishing Group}
}

@article{lyubarov2022amplified,
  title={Amplified emission and lasing in photonic time crystals},
  author={Lyubarov, Mark and Lumer, Yaakov and Dikopoltsev, Alex and Lustig, Eran and Sharabi, Yonatan and Segev, Mordechai},
  journal={Science},
  volume={377},
  number={6604},
  pages={425--428},
  year={2022},
  publisher={American Association for the Advancement of Science}
}

@article{galiffi2026optical,
  title={Optical coherent perfect absorption and amplification in a time-varying medium},
  author={Galiffi, Emanuele and Harwood, Anthony C and Vezzoli, Stefano and Tirole, Romain and Al{\`u}, Andrea and Sapienza, Riccardo},
  journal={Nature Photonics},
  pages={1--7},
  year={2026},
  publisher={Nature Publishing Group UK London}
}

@article{galiffi2025electrodynamics,
  title={Electrodynamics of photonic temporal interfaces},
  author={Galiffi, Emanuele and Sol{\'\i}s, Diego Mart{\'\i}nez and Yin, Shixiong and Engheta, Nader and Al{\`u}, Andrea},
  journal={Light: Science \& Applications},
  volume={14},
  number={1},
  pages={338},
  year={2025},
  publisher={Nature Publishing Group UK London}
}

@article{salary2019dynamically,
  title={A dynamically modulated all-dielectric metasurface doublet for directional harmonic generation and manipulation in transmission},
  author={Salary, Mohammad Mahdi and Farazi, Soheil and Mosallaei, Hossein},
  journal={Advanced Optical Materials},
  volume={7},
  number={23},
  pages={1900843},
  year={2019},
  publisher={Wiley Online Library}
}

@article{shaltout2015time,
  title={Time-varying metasurfaces and Lorentz non-reciprocity},
  author={Shaltout, Amr and Kildishev, Alexander and Shalaev, Vladimir},
  journal={Optical Materials Express},
  volume={5},
  number={11},
  pages={2459--2467},
  year={2015},
  publisher={Optical Society of America}
}

@article{tirole2023double,
  title={Double-slit time diffraction at optical frequencies},
  author={Tirole, Romain and Vezzoli, Stefano and Galiffi, Emanuele and Robertson, Iain and Maurice, Dries and Tilmann, Benjamin and Maier, Stefan A and Pendry, John B and Sapienza, Riccardo},
  journal={Nature Physics},
  volume={19},
  number={7},
  pages={999--1002},
  year={2023},
  publisher={Nature Publishing Group UK London}
}

@article{dong2024quantum,
  title={Quantum time reflection and refraction of ultracold atoms},
  author={Dong, Zhaoli and Li, Hang and Wan, Tuo and Liang, Qian and Yang, Zhaoju and Yan, Bo},
  journal={Nature Photonics},
  volume={18},
  number={1},
  pages={68--73},
  year={2024},
  publisher={Nature Publishing Group UK London}
}

@article{koutserimpas2018parametric,
  title={Parametric amplification and bidirectional invisibility in PT-symmetric time-Floquet systems},
  author={Koutserimpas, Theodoros T and Al{\`u}, Andrea and Fleury, Romain},
  journal={Physical Review A},
  volume={97},
  number={1},
  pages={013839},
  year={2018},
  publisher={APS}
}

@article{sadafi2023dynamic,
  title={Dynamic control of light scattering in a single particle enabled by time modulation},
  author={Sadafi, Mohammad Mojtaba and da Mota, Achiles Fontana and Mosallaei, Hossein},
  journal={Applied Physics Letters},
  volume={123},
  number={10},
  pages={101702},
  year={2023},
  publisher={AIP Publishing LLC}
}

@article{asadchy2022parametric,
  title={Parametric Mie resonances and directional amplification in time-modulated scatterers},
  author={Asadchy, V and Lamprianidis, AG and Ptitcyn, G and Albooyeh, M and Rituraj and Karamanos, T and Alaee, R and Tretyakov, SA and Rockstuhl, C and Fan, S},
  journal={Physical Review Applied},
  volume={18},
  number={5},
  pages={054065},
  year={2022},
  publisher={APS}
}

@article{estep2014magnetic,
  title={Magnetic-free non-reciprocity and isolation based on parametrically modulated coupled-resonator loops},
  author={Estep, Nicholas A and Sounas, Dimitrios L and Soric, Jason and Alu, Andrea},
  journal={Nature Physics},
  volume={10},
  number={12},
  pages={923--927},
  year={2014},
  publisher={Nature Publishing Group UK London}
}

@article{barati2022optical,
  title={Optical nonreciprocity via transmissive time-modulated metasurfaces},
  author={Barati Sedeh, Hooman and Mohammadi Dinani, Hediyeh and Mosallaei, Hossein},
  journal={Nanophotonics},
  volume={11},
  number={17},
  pages={4135--4148},
  year={2022},
  publisher={De Gruyter}
}

@article{wang2024time,
  title={Time-varying systems to improve the efficiency of wireless power transfer},
  author={Wang, Xuchen and Krois, I and Ha-Van, N and Mirmoosa, MS and Jayathurathnage, P and Hrabar, S and Tretyakov, SA},
  journal={Physical review applied},
  volume={21},
  number={5},
  pages={054027},
  year={2024},
  publisher={APS}
}

@article{sadafi2025time,
  title={Time-varying Mie resonators for real-time manipulation of quantum emitter radiation},
  author={Sadafi, Mohammad Mojtaba and da Mota, Achiles Fontana and Mosallaei, Hossein},
  journal={Physical Review B},
  volume={111},
  number={12},
  pages={125419},
  year={2025},
  publisher={APS}
}

@article{da2025dynamics,
  title={Dynamics of time-modulated quantum systems via integrated Lindblad and Maxwell--Bloch equations},
  author={da Mota, Achiles F and Sadafi, Mohammad Mojtaba and Chiu, Wei-Chi and Barbiellini, Bernardo and Leuenberger, Michael N and Bansil, Arun and Mosallaei, Hossein},
  journal={APL Quantum},
  volume={2},
  number={3},
  year={2025},
  publisher={AIP Publishing}
}

@article{horsley2023quantum,
  title={Quantum electrodynamics of time-varying gratings},
  author={Horsley, Simon AR and Pendry, John B},
  journal={Proceedings of the National Academy of Sciences},
  volume={120},
  number={36},
  pages={e2302652120},
  year={2023},
  publisher={National Academy of Sciences}
}

@article{zhu2022time,
  title={Time-periodic corner states from Floquet higher-order topology},
  author={Zhu, Weiwei and Xue, Haoran and Gong, Jiangbin and Chong, Yidong and Zhang, Baile},
  journal={Nature Communications},
  volume={13},
  number={1},
  pages={11},
  year={2022},
  publisher={Nature Publishing Group UK London}
}

@article{xiong2025observation,
  title={Observation of wave amplification and temporal topological state in a non-synthetic photonic time crystal},
  author={Xiong, Jiang and Zhang, Xudong and Duan, Longji and Wang, Jiarui and Long, Yang and Hou, Haonan and Yu, Letian and Zou, Linyang and Zhang, Baile},
  journal={Nature Communications},
  volume={16},
  number={1},
  pages={11182},
  year={2025},
  publisher={Nature Publishing Group UK London}
}

@article{dikopoltsev2022light,
  title={Light emission by free electrons in photonic time-crystals},
  author={Dikopoltsev, Alex and Sharabi, Yonatan and Lyubarov, Mark and Lumer, Yaakov and Tsesses, Shai and Lustig, Eran and Kaminer, Ido and Segev, Mordechai},
  journal={Proceedings of the National Academy of Sciences},
  volume={119},
  number={6},
  pages={e2119705119},
  year={2022},
  publisher={National Academy of Sciences}
}

@article{li2023stationary,
  title={Stationary charge radiation in anisotropic photonic time crystals},
  author={Li, Huanan and Yin, Shixiong and He, Huan and Xu, Jingjun and Al{\`u}, Andrea and Shapiro, Boris},
  journal={Physical Review Letters},
  volume={130},
  number={9},
  pages={093803},
  year={2023},
  publisher={APS}
}

@article{sabri2023high,
  title={High-quality-factor space--time metasurface for free-space power isolation at near-infrared regime},
  author={Sabri, Raana and Mosallaei, Hossein},
  journal={Advanced Photonics Nexus},
  volume={2},
  number={6},
  pages={066008--066008},
  year={2023},
  publisher={Society of Photo-Optical Instrumentation Engineers}
}

@article{barati2020topological,
  title={Topological space-time photonic transitions in angular-momentum-biased metasurfaces},
  author={Barati Sedeh, Hooman and Salary, Mohammad Mahdi and Mosallaei, Hossein},
  journal={Advanced Optical Materials},
  volume={8},
  number={11},
  pages={2000075},
  year={2020},
  publisher={Wiley Online Library}
}

@article{wu2025space,
  title={A space-time holographic metasurface antenna},
  author={Wu, Geng-Bo and Dai, Jun Yan and Sun, Yiqing and Shum, Kam Man and Chan, Ka Fai and Cheng, Qiang and Cui, Tie Jun and Chan, Chi Hou},
  journal={Science Advances},
  volume={11},
  number={38},
  pages={eadx7090},
  year={2025},
  publisher={American Association for the Advancement of Science}
}

@article{hu2022arbitrary,
  title={Arbitrary and dynamic Poincar{\'e} sphere polarization converter with a time-varying metasurface},
  author={Hu, Qi and Chen, Ke and Zhang, Na and Zhao, Jianming and Jiang, Tian and Zhao, Junming and Feng, Yijun},
  journal={Advanced Optical Materials},
  volume={10},
  number={4},
  pages={2101915},
  year={2022},
  publisher={Wiley Online Library}
}

@article{garg2025inverse,
  title={Inverse-Designed Dispersive Time-Varying Nanostructures},
  author={Garg, Puneet and Fischbach, Jan David and Lamprianidis, Aristeidis G and Wang, Xuchen and Mirmoosa, Mohammad S and Asadchy, Viktar S and Rockstuhl, Carsten and Sturges, Thomas J},
  journal={Advanced Optical Materials},
  volume={13},
  number={5},
  pages={2402444},
  year={2025},
  publisher={Wiley Online Library}
}

@article{barati2020time,
  title={Time-varying optical vortices enabled by time-modulated metasurfaces},
  author={Barati Sedeh, Hooman and Salary, Mohammad Mahdi and Mosallaei, Hossein},
  journal={Nanophotonics},
  volume={9},
  number={9},
  pages={2957--2976},
  year={2020},
  publisher={De Gruyter}
}

@article{wang2025expanding,
  title={Expanding momentum bandgaps in photonic time crystals through resonances},
  author={Wang, X and Garg, P and Mirmoosa, MS and Lamprianidis, AG and Rockstuhl, C and Asadchy, VS},
  journal={Nature Photonics},
  volume={19},
  number={2},
  pages={149--155},
  year={2025},
  publisher={Nature Publishing Group UK London}
}

@article{ma2025floquet,
  title={Floquet topological states in time-varying metasurfaces},
  author={Ma, Qian and You, Jian Wei and Chen, Long and Gu, Ze and Qin, Shi Long and Lan, Zhihao and Cui, Tie Jun},
  journal={Science Advances},
  volume={11},
  number={39},
  pages={eadx9025},
  year={2025},
  publisher={American Association for the Advancement of Science}
}

@article{sisler2024electrically,
  title={Electrically tunable space--time metasurfaces at optical frequencies},
  author={Sisler, Jared and Thureja, Prachi and Grajower, Meir Y and Sokhoyan, Ruzan and Huang, Ivy and Atwater, Harry A},
  journal={Nature Nanotechnology},
  volume={19},
  number={10},
  pages={1491--1498},
  year={2024},
  publisher={Nature Publishing Group UK London}
}

@article{jaffray2026all,
  title={All-optical polarization control in time-varying low-index films via plasma symmetry breaking},
  author={Jaffray, Wallace and Stengel, Sven and Boltasseva, Alexandra and Shalaev, Vladimir M and Vincenti, Maria Antonietta and De Ceglia, Domenico and Scalora, Michael and Rizza, Carlo and Ferrera, Marcello},
  journal={Nature Photonics},
  pages={1--9},
  year={2026},
  publisher={Nature Publishing Group UK London}
}

@article{liu2021photon,
  title={Photon acceleration using a time-varying epsilon-near-zero metasurface},
  author={Liu, Cong and Alam, M Zahirul and Pang, Kai and Manukyan, Karapet and Reshef, Orad and Zhou, Yiyu and Choudhary, Saumya and Patrow, Joel and Pennathurs, Anuj and Song, Hao and others},
  journal={Acs Photonics},
  volume={8},
  number={3},
  pages={716--720},
  year={2021},
  publisher={ACS Publications}
}

@article{dhama2026cross,
  title={Cross-Phase Modulation via Time-Varying Epsilon-Near-Zero Metasurfaces},
  author={Dhama, Rakesh and Hossain, Md Imran and Pietila, Jesse and Fordell, Thomas and Caglayan, Humeyra},
  journal={Advanced Optical Materials},
  pages={e03406},
  year={2026},
  publisher={Wiley Online Library}
}

@article{kumar2021light,
  title={Light-Matter Coupling in Scalable Van der Waals Superlattices},
  author={Kumar, Pawan and Lynch, Jason and Song, Baokun and Ling, Haonan and Barrera, Francisco and Zhang, Huiqin and Anantharaman, Surendra B and Digani, Jagrit and Zhu, Haoyue and Choudhury, Tanushree H and others},
  journal={arXiv preprint arXiv:2103.14028},
  year={2021}
}

@article{lynch2025electrically,
  title={Electrically Tunable Excitonic-Hyperbolicity in Chirality-Pure Carbon Nanotubes},
  author={Lynch, Jason and Shapturenka, Pavel and Sadafi, Mohammad Mojtaba and Liu, Zoey and Ruth, Tobia and Jha, Kritika and Fakhraai, Zahra and Mosallaei, Hossein and Engheta, Nader and Fagan, Jeffrey A and others},
  journal={arXiv preprint arXiv:2509.24848},
  year={2025}
}

@article{di2025efficient,
  title={Efficient GHz electro-optical modulation with a nonlocal lithium niobate metasurface in the linear and nonlinear regime},
  author={Di Francescantonio, Agostino and Sabatti, Alessandra and Weigand, Helena and Bailly-Rioufreyt, Elise and Vincenti, Maria Antonietta and Carletti, Luca and Kellner, Jost and Zilli, Attilio and Finazzi, Marco and Celebrano, Michele and others},
  journal={Nature Communications},
  volume={16},
  number={1},
  pages={7000},
  year={2025},
  publisher={Nature Publishing Group UK London}
}

@article{wang2018integrated,
  title={Integrated lithium niobate electro-optic modulators operating at CMOS-compatible voltages},
  author={Wang, Cheng and Zhang, Mian and Chen, Xi and Bertrand, Maxime and Shams-Ansari, Amirhassan and Chandrasekhar, Sethumadhavan and Winzer, Peter and Lon{\v{c}}ar, Marko},
  journal={Nature},
  volume={562},
  number={7725},
  pages={101--104},
  year={2018},
  publisher={Nature Publishing Group UK London}
}

@article{lynch2025full,
  title={Full 2$\pi$ phase modulation using exciton-polaritons in a two-dimensional superlattice},
  author={Lynch, Jason and Kumar, Pawan and Chen, Chen and Trainor, Nicholas and Kumari, Shalina and Peng, Tzu-Yu and Chen, Cindy Yueli and Lu, Yu-Jung and Redwing, Joan and Jariwala, Deep},
  journal={Device},
  volume={3},
  number={1},
  year={2025},
  publisher={Elsevier}
}

@article{sadafi2021tunable,
  title={A tunable hybrid graphene-metal metamaterial absorber for sensing in the THz regime},
  author={Sadafi, Mohammad Mojtaba and Karami, Hamidreza and Hosseini, Manouchehr},
  journal={Current Applied Physics},
  volume={31},
  pages={132--140},
  year={2021},
  publisher={Elsevier}
}

@article{khurgin2024energy,
  title={Energy and power requirements for alteration of the refractive index},
  author={Khurgin, Jacob B},
  journal={Laser \& Photonics Reviews},
  volume={18},
  number={4},
  pages={2300836},
  year={2024},
  publisher={Wiley Online Library}
}

@article{hayran2022homega,
  title={{$\hbar\omega$ versus $\hbar k$: dispersion and energy constraints on time-varying photonic materials and time crystals}},
  author={Hayran, Zeki and Khurgin, Jacob B. and Monticone, Francesco},
  journal={Optical Materials Express},
  volume={12},
  number={10},
  pages={3904--3917},
  year={2022},
  publisher={Optica Publishing Group}
}

@article{sloan2024optical,
  title={Optical properties of dispersive time-dependent materials},
  author={Sloan, Jamison and Rivera, Nicholas and Joannopoulos, John D and Soljacic, Marin},
  journal={ACS Photonics},
  volume={11},
  number={3},
  pages={950--962},
  year={2024},
  publisher={ACS Publications}
}

@article{koutserimpas2024time,
  title={Time-varying media, dispersion, and the principle of causality},
  author={Koutserimpas, Theodoros T and Monticone, Francesco},
  journal={Optical Materials Express},
  volume={14},
  number={5},
  pages={1222--1236},
  year={2024},
  publisher={Optica Publishing Group}
}

@article{salary2018time,
  title={Time-varying metamaterials based on graphene-wrapped microwires: Modeling and potential applications},
  author={Salary, Mohammad Mahdi and Jafar-Zanjani, Samad and Mosallaei, Hossein},
  journal={Physical Review B},
  volume={97},
  number={11},
  pages={115421},
  year={2018},
  publisher={APS}
}

@article{inampudi2018rigorous,
  title={Rigorous space-time coupled-wave analysis for patterned surfaces with temporal permittivity modulation},
  author={Inampudi, Sandeep and Salary, Mohammad Mahdi and Jafar-Zanjani, Samad and Mosallaei, Hossein},
  journal={Optical Materials Express},
  volume={9},
  number={1},
  pages={162--182},
  year={2018},
  publisher={Optical Society of America}
}

@article{stefanou2023light,
  title={Light scattering by a periodically time-modulated object of arbitrary shape: the extended boundary condition method},
  author={Stefanou, Nikolaos and Stefanou, Ioannis and Almpanis, Evangelos and Papanikolaou, Nikolaos and Garg, Puneet and Rockstuhl, Carsten},
  journal={Journal of the Optical Society of America B},
  volume={40},
  number={11},
  pages={2842--2850},
  year={2023},
  publisher={Optica Publishing Group}
}

@article{taravati2017nonreciprocal,
  title={Nonreciprocal electromagnetic scattering from a periodically space-time modulated slab and application to a quasisonic isolator},
  author={Taravati, Sajjad and Chamanara, Nima and Caloz, Christophe},
  journal={Physical Review B},
  volume={96},
  number={16},
  pages={165144},
  year={2017},
  publisher={APS}
}

@article{mirmoosa2022dipole,
  title={Dipole polarizability of time-varying particles},
  author={Mirmoosa, Mohammad Sajjad and Koutserimpas, TT and Ptitcyn, GA and Tretyakov, SA and Fleury, R},
  journal={New Journal of Physics},
  volume={24},
  number={6},
  pages={063004},
  year={2022},
  publisher={IOP Publishing}
}

@article{ptitcyn2023floquet,
  title={Floquet--Mie theory for time-varying dispersive spheres},
  author={Ptitcyn, Grigorii and Lamprianidis, Aristeidis and Karamanos, Theodosios and Asadchy, Viktar and Alaee, Rasoul and M{\"u}ller, Marvin and Albooyeh, Mohammad and Mirmoosa, Mohammad Sajjad and Fan, Shanhui and Tretyakov, Sergei and others},
  journal={Laser \& Photonics Reviews},
  volume={17},
  number={3},
  pages={2100683},
  year={2023},
  publisher={Wiley Online Library}
}

@article{garg2022modeling,
  title={Modeling four-dimensional metamaterials: a T-matrix approach to describe time-varying metasurfaces},
  author={Garg, Puneet and Lamprianidis, Aristeidis G and Beutel, Dominik and Karamanos, Theodosios and Verf{\"u}rth, Barbara and Rockstuhl, Carsten},
  journal={Optics Express},
  volume={30},
  number={25},
  pages={45832--45847},
  year={2022},
  publisher={Optica Publishing Group}
}

@article{sun2025formulation,
  title={Formulation of dispersive and dissipative time-varying media as a Floquet matrix eigenproblem},
  author={Sun, Yuchen and Fan, Shanhui and Hu, Guangwei},
  journal={Physical Review Letters},
  volume={135},
  number={15},
  pages={156903},
  year={2025},
  publisher={APS}
}

@article{de2025lattice,
  title={Lattice Resonances in Periodic Arrays of Time-Modulated Scatterers},
  author={de Paz, Mar{\'\i}a Blanco and Deop-Ruano, Juan R and Sol{\'\i}s, Diego M and Manjavacas, Alejandro},
  journal={arXiv preprint arXiv:2511.11454},
  year={2025}
}

@article{verde2026optical,
  title={Optical response by time-varying plasmonic nanoparticles},
  author={Verde, Miguel and Huidobro, Paloma A},
  journal={Physical Review Research},
  volume={8},
  number={2},
  pages={023073},
  year={2026},
  publisher={APS}
}

@article{iplikcciouglu2025analytical,
  title={Analytical modeling of time-varying and dispersive metasurfaces with surface susceptibility operators},
  author={{\.I}plik{\c{c}}io{\u{g}}lu, Suat Bar{\i}{\c{s}} and Aksun, MI},
  journal={Physical Review B},
  volume={111},
  number={24},
  pages={245141},
  year={2025},
  publisher={APS}
}

@article{movahediqomi2026stacked,
  title={Stacked Time-Varying Metasurfaces},
  author={Movahediqomi, Mostafa and Tretyakov, Sergei and Asadchy, Viktar and Wang, Xuchen},
  journal={Advanced Optical Materials},
  year={2026},
  publisher={Wiley}
}

@article{moharam1981rigorous,
  title   = {Rigorous coupled-wave analysis of planar-grating diffraction},
  author  = {Moharam, M. G. and Gaylord, T. K.},
  journal = {Journal of the Optical Society of America},
  volume  = {71},
  number  = {7},
  pages   = {811--818},
  year    = {1981},
  doi     = {10.1364/JOSA.71.000811}
}

@article{moharam1995stable,
  title   = {Stable implementation of the rigorous coupled-wave analysis for surface-relief gratings: enhanced transmittance matrix approach},
  author  = {Moharam, M. G. and Pommet, Drew A. and Grann, Eric B. and Gaylord, T. K.},
  journal = {Journal of the Optical Society of America A},
  volume  = {12},
  number  = {5},
  pages   = {1077--1086},
  year    = {1995},
  doi     = {10.1364/JOSAA.12.001077}
}

@book{stratton2007electromagnetic,
  title={Electromagnetic theory},
  author={Stratton, Julius Adams},
  year={2007},
  publisher={John Wiley \& Sons}
}

@misc{yeh1990optical,
  title={Optical waves in layered media},
  author={Yeh, Pochi and Hendry, Michael},
  year={1990},
  publisher={American Institute of Physics}
}

@book{fox2010optical,
  title={Optical properties of solids},
  author={Fox, Mark},
  volume={3},
  year={2010},
  publisher={Oxford university press}
}

@article{cong2018optical,
  title={Optical properties of 2D semiconductor WS2},
  author={Cong, Chunxiao and Shang, Jingzhi and Wang, Yanlong and Yu, Ting},
  journal={Advanced Optical Materials},
  volume={6},
  number={1},
  pages={1700767},
  year={2018},
  publisher={Wiley Online Library}
}

@article{bianchi2024engineering,
  title={Engineering the electrical and optical properties of WS2 monolayers via defect control},
  author={Bianchi, Michele Giovanni and Risplendi, Francesca and Re Fiorentin, Michele and Cicero, Giancarlo},
  journal={Advanced Science},
  volume={11},
  number={4},
  pages={2305162},
  year={2024},
  publisher={Wiley Online Library}
}

@article{li2014measurement,
  title={Measurement of the optical dielectric function of monolayer transition-metal dichalcogenides: MoS 2, Mo S e 2, WS 2, and WS e 2},
  author={Li, Yilei and Chernikov, Alexey and Zhang, Xian and Rigosi, Albert and Hill, Heather M and Van Der Zande, Arend M and Chenet, Daniel A and Shih, En-Min and Hone, James and Heinz, Tony F},
  journal={Physical Review B},
  volume={90},
  number={20},
  pages={205422},
  year={2014},
  publisher={APS}
}

@article{weber2023intrinsic,
  title={Intrinsic strong light-matter coupling with self-hybridized bound states in the continuum in van der Waals metasurfaces},
  author={Weber, Thomas and K{\"u}hner, Lucca and Sortino, Luca and Ben Mhenni, Amine and Wilson, Nathan P and K{\"u}hne, Julius and Finley, Jonathan J and Maier, Stefan A and Tittl, Andreas},
  journal={Nature Materials},
  volume={22},
  number={8},
  pages={970--976},
  year={2023},
  publisher={Nature Publishing Group UK London}
}

@incollection{sie2017valley,
  title={Valley-selective optical Stark effect in monolayer WS2},
  author={Sie, Edbert Jarvis},
  booktitle={Coherent Light-Matter Interactions in Monolayer Transition-Metal Dichalcogenides},
  pages={37--57},
  year={2017},
  publisher={Springer}
}

@article{cunningham2019resonant,
  title={Resonant optical Stark effect in monolayer WS2},
  author={Cunningham, Paul D and Hanbicki, Aubrey T and Reinecke, Thomas L and McCreary, Kathleen M and Jonker, Berend T},
  journal={Nature communications},
  volume={10},
  number={1},
  pages={5539},
  year={2019},
  publisher={Nature Publishing Group UK London}
}

@article{chakraborty2018control,
  title={Control of strong light--matter interaction in monolayer WS2 through electric field gating},
  author={Chakraborty, Biswanath and Gu, Jie and Sun, Zheng and Khatoniar, Mandeep and Bushati, Rezlind and Boehmke, Alexandra L and Koots, Rian and Menon, Vinod M},
  journal={Nano letters},
  volume={18},
  number={10},
  pages={6455--6460},
  year={2018},
  publisher={ACS Publications}
}

@article{he2016strain,
  title={Strain engineering in monolayer WS2, MoS2, and the WS2/MoS2 heterostructure},
  author={He, Xin and Li, Hai and Zhu, Zhiyong and Dai, Zhenyu and Yang, Yang and Yang, Peng and Zhang, Qiang and Li, Peng and Schwingenschlogl, Udo and Zhang, Xixiang},
  journal={Applied Physics Letters},
  volume={109},
  number={17},
  year={2016},
  publisher={AIP Publishing}
}

@article{boltasseva2024photonic,
  title={Photonic time crystals: from fundamental insights to novel applications: opinion},
  author={Boltasseva, A and Shalaev, VM and Segev, M},
  journal={Optical Materials Express},
  volume={14},
  number={3},
  pages={592--597},
  year={2024},
  publisher={Optica Publishing Group}
}

@article{garg2025photonic,
  title={Photonic time crystals assisted by quasi-bound states in the continuum},
  author={Garg, P and Almpanis, E and Zimmer, L and Fischbach, JD and Wang, X and Mirmoosa, MS and Nyman, M and Stefanou, N and Papanikolaou, N and Asadchy, V and others},
  journal={arXiv preprint arXiv:2507.15644},
  year={2025}
}

@article{huang2026observation,
  title={Observation of full momentum bandgap in photonic time crystals},
  author={Huang, Bolun and Zhu, Zebin and Yu, Genrong and Gao, Zhen},
  journal={arXiv preprint arXiv:2604.17408},
  year={2026}
}
\bibliographystyle{ieeetr}

\end{document}